\documentclass[aps,pre,twocolumn,groupedaddress,floatfix,longbibliography,superscriptaddress]{revtex4-1}

\usepackage{amsmath}
\usepackage{amssymb}
\usepackage{amsfonts}
\usepackage{graphicx}
\usepackage{bbold}
\usepackage{bm}
\usepackage{bm}
\usepackage{cancel}
\usepackage{times,float}
\usepackage{graphicx}
\usepackage[usenames,dvipsnames,svgnames]{xcolor}
\usepackage{hyperref}
\hypersetup{colorlinks=true, linkcolor=Blue, citecolor=Blue,urlcolor=Blue}
\usepackage{multirow}
\usepackage{ulem}
\usepackage{color,epsfig}
\usepackage{bm}
\usepackage{slashed}
\usepackage{hyperref}
\usepackage{subfigure}
\usepackage{feynmf}
\usepackage{mathrsfs}
\usepackage{simpler-wick}
\usepackage{orcidlink}

\newcommand{\bea}{\begin{eqnarray}}
\newcommand{\eea}{\end{eqnarray}}
\newcommand{\be}{\begin{equation}}
\newcommand{\ee}{\end{equation}}

\newcommand{\sgn}{{\rm sign}}

\newcommand{\nn}{\nonumber}
\newcommand{\ii}{\mathrm{i}}

\newcommand{\qB}{|q_fB|}
\newcommand{\kt}{\mathbf{k}_\perp}
\newcommand{\pt}{\mathbf{p}_\perp}
\newcommand{\sign}{\text{sign}(q_fB)}

\newcommand{\kp}{k_\parallel}

\newcommand{\gn}{\gamma^\nu}
\newcommand{\gm}{\gamma^\mu}

\newcommand{\pp}{p_\parallel}

\newcommand{\Op}[1]{\mathcal{O}^{(#1)}}

\newcommand{\wn}{\omega_n}

\begin{document}

\title{Anisotropic Photon and Dilepton Yield in a Thermalized Quark-Gluon Plasma under Magnetic Fluctuations}
\author{Jorge David Casta\~no-Yepes~\orcidlink{0000-0002-8654-1304}}
\email{jcastano@uc.cl}
\affiliation{Facultad de F\'isica, Pontificia Universidad Cat\'olica de Chile, Vicu\~{n}a Mackenna 4860, Santiago, Chile}
\author{Enrique Mu\~noz~\orcidlink{0000-0003-4457-0817}}
\email{ejmunozt@uc.cl}
\affiliation{Facultad de F\'isica, Pontificia Universidad Cat\'olica de Chile, Vicu\~{n}a Mackenna 4860, Santiago, Chile}
\affiliation{Center for Nanotechnology and Advanced Materials CIEN-UC, Avenida Vicuña Mackenna 4860, Santiago, Chile}

\begin{abstract}
In this article, we analyze the effects of stochastic magnetic fluctuations with respect to an intense magnetic field background over the yields for photon and dilepton emission processes in a thermalized quark-gluon plasma phase. Such stochastic fluctuations model the effects of nearly random initial conditions for the nuclei participating in non-central heavy-ion collisions, which are the sources of the background magnetic field. Our theoretical results predict significant anisotropic effects due to stochastic magnetic noise over the angular distribution for photon and dilepton production rates in this scenario.
\end{abstract}

\maketitle 

\section{Introduction}

\begin{figure}
    \centering
    \includegraphics[scale=1]{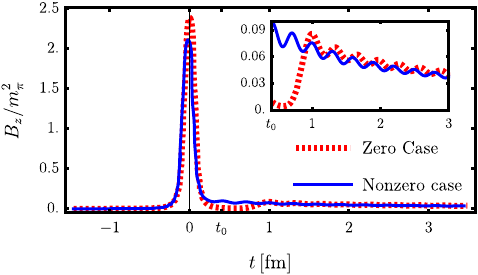}
    \caption{Magnetic field with different models for QCD electric conductivities. Taken from Ref.~\cite{PhysRevC.107.034901} (``nonzero case'' means that quarks in the QGP are created in the time window $0<t\leq t_0$, with $t_0=0.4$ fm).}
    \label{fig:DataPhysRevC107_034901_2023}
\end{figure}

Current experiments on ultra-relativistic heavy-ion collisions (HIC) constitute our state-of-the-art methodology to probe the properties of strongly correlated quantum matter, at high energy density and pressure, as examined under the rationale of Quantum Chromodynamics (QCD). Even though thermalization occurs quite early in the resulting deconfined quark-gluon plasma (QGP) phase, strong transient effects depending on nearly random initial conditions prior to the collision emerge. Among those, the presence of very intense, however, fastly decaying magnetic fields (see Fig.~\ref{fig:DataPhysRevC107_034901_2023} with data from Ref.~\cite{PhysRevC.107.034901}) arising from the initial spatial asymmetry in the overlap region of nuclei in non-central collisions, is a subject of major concern for the interpretation of the experimental signals detected and the corresponding physical properties of the resulting magnetized medium. It has been argued that, given that the sources of the background magnetic field are the charged nucleons, its characteristic decay time is determined by the effective conductivity of the medium~\cite{PhysRevC.107.034901,Hattori_2016,Mizher_PhysRevD.110.L111501}. In this scenario, it is clear that the simplifying assumption of a stationary and spatially uniform magnetic field background is quite far from reality and it should be improved in order to better interpret the experimental signals in HIC.

In an attempt to go beyond the standard approximation of a constant magnetic field, usually incorporated into the fermion propagators after the seminal work by Schwinger~\cite{Schwinger_1951,Dittrich_Reuter}, we recently developed a theory that includes white-noise-correlated stochastic fluctuations over an otherwise uniform average classical background field~\cite{PhysRevD.107.096014,PhysRevD.108.116013}. For the QED gauge fields, we distinguish three physically different contributions
\begin{eqnarray}
A^{\mu}(x) + A^{\mu}_\text{BG}(x) + \delta A^{\mu}_\text{BG}(\mathbf{x}),
\label{eq_Atot}
\end{eqnarray}
where $A^{\mu}(x)$ is the quantum dynamical gauge field (photons), while BG represents a classical ``background''. In addition, we introduce classical white noise fluctuations $\delta A^{\mu}_\text{BG}(\mathbf{x})$ with respect to the mean value $A_\text{BG}^{\mu}(x)$, with the following statistical properties
\begin{eqnarray}
\langle \delta A^{j}_\text{BG}(\mathbf{x}) \delta A^{k}_\text{BG}(\mathbf{x}')\rangle &=&  \Delta\delta_{j,k}\delta^{(3)}(\mathbf{x}-\mathbf{x}'),\nonumber\\
\langle \delta A^{\mu}_\text{BG}(\mathbf{x})\rangle &=& 0.
\label{eq_Acorr}
\end{eqnarray}

As a consequence, the classical magnetic field background inherits those stochastic fluctuations $\mathbf{B} + \delta\mathbf{B}(x) = \nabla\times\left( \mathbf{A}_\text{BG}(x) + \delta \mathbf{A}_\text{BG}(x) \right)$, with a uniform mean value $\mathbf{B}$.

In our theory, we perform a statistical average over a functional distribution of the random fluctuations $\delta \mathbf{A}_\text{BG}(x)$ of the classical background by means of the replica method~\cite{Parisi_Mezard_1991,Kardar_Parisi_1986}. We then obtained explicit analytical expressions for the fermion~\cite{PhysRevD.107.096014} and gauge field propagators~\cite{PhysRevD.109.056007}, respectively, as a perturbative expansion in terms of the magnetic noise autocorrelation $\Delta \ge 0$. As a consequence, a quasi-particle interpretation of the noise effects emerges~\cite{PhysRevD.109.056007,PhysRevD.107.096014}, thus leading to a renormalized magnetic mass for the fermions~\cite{PhysRevD.110.056003}. Moreover, as we recently showed, classical magnetic fluctuations combined with charged vacuum polarization in QED generate anisotropic photon magnetic masses $M_{\perp}$ and $M_{\parallel}$ which are proportional to $\Delta$~, such that the physical picture is that photons propagate in a birefringent, magnetized and dispersive effective medium~\cite{PhysRevD.109.056007}. 

An important physical question that then arises is the possibility for those highly non-trivial effects to manifest themselves in the experimental signals detected in HIC. 
In particular, electromagnetic probes such as direct photons and real dilepton pairs ($e^+ e^-$ and $\mu^+ \mu^-$) do not possess color charges, thus being immune to strong interactions emerging in the final states of QCD matter~\cite{Ryblewsky_PhysRevD.92.025026,SHEN2016184,Monnai_2020}. Therefore, they provide valuable information about the microscopic processes from which they originate. Regarding inclusive photons, they consist on direct photons generated by primary sources, and decay photons produced by the subsequent hadronic decay processes.

Direct photons~\cite{Monnai_2020,SHEN2016184} are commonly estimated as the sum of prompt photons~\cite{PhysRevD.96.014023,Ayala2020,PhysRevC.106.064905} produced in the hard process at the collision instant, plus thermal photons emitted from the soft, thermalized portion of the medium. This is an oversimplification, since nuclear matter undergoes several stages after the collision, from initial color glass condensate to glasma, then a hydrodynamic QCD regime, and finally hadronic gas. Therefore, both pre-equilibrium and post-hydrodynamic stages should be distinguished for a more accurate description of direct photons.
Whereas the commonly accepted picture is that the QGP behaves as a nearly perfect hydrodynamic fluid, azimuthal asymmetries in photon production rates have been experimentally detected in a wide range of rapidities by the PHENIX collaboration at the RHIC~\cite{Adare_PhysRevLett.109.122302,Adare_PhysRevC.94.064901}, and by the ALICE collaboration at the LHC~\cite{ALICE_2019308}.
Those effects are characterized by the anisotropic flow coefficients $\left\{v_n\right\}_{n=1}^{\infty}$ arising from the Fourier decomposition of the azimuthal angular distribution of particles~\cite{PhysRevD.102.076010,PhysRevD.109.056008}, 
\bea
p^0\frac{d^3 R}{dp_x dp_y dp_z} &=& \frac{d^3 R}{p_T dp_T d\phi d y}\nn\\
&=&\frac{1}{2\pi} \mathcal{R}_0\left[ 1 + 2 \sum_{n=1}^{\infty}v_n\cos(n\phi)\right].
\label{ec:defflujos}
\eea

Here $\phi$ represents the angle between the emitted particles (photons) and the reaction plane, such that $p^{\mu} = (p^0,\mathbf{p})$ is resolved into parallel and transverse components $p_T = \sqrt{p_y^2 + p_z^2}$, with
\be
p_y = p_T\cos\phi,\,\,\,p_z=p_T\sin\phi,
\ee
while $y = \frac{1}{2}\ln\frac{p_0 + p_x}{p_0-p_x}$ represents the rapidity. 
It follows directly from this definition that the anisotropy flow coefficients $v_n$ are defined by the integral expressions
\bea
v_n = \frac{1}{\mathcal{R}_0}\int_0^{2\pi}d\phi\cos(n\phi)\frac{d^3 R}{p_T dp_T d\phi d y},
\label{eq:vn}
\eea
with
\bea
\mathcal{R}_0 = \frac{1}{2\pi}\int_0^{2\pi}d\phi\frac{d^3 R}{p_T dp_T d\phi d y}.
\eea

The direct photon emission rate is known to exhibit the ``photon puzzle"~\cite{PhysRevC.93.044906,SHEN2016184,PhysRevC.98.054902,Monnai_2020, PhysRevC.107.024914}, where the azimuthal momentum anisotropy characterized by the magnitude of the elliptic and triangular flow coefficients is larger than predicted from hydrodynamic calculations.
It can be shown that the field-theoretical expression for the photon emission rate distribution by pair annihilation depends on the imaginary part of the retarded polarization tensor $\Pi_{\text{R}}^{\mu\nu}(p)$, according to the formula~\cite{PhysRevD.110.056003,PhysRevD.109.056008}
\bea
\frac{d^3R}{p_\perp dp_\perp d\phi dy}=-\frac{n_{\text{B}}(\omega)}{(2\pi)^3}\text{Im}\left\{g_{\mu\nu}\Pi_{\text{R}}^{\mu\nu}(p)\right\},
\label{eq:Rgamma}
\eea
where $n_{\text{B}}(\omega) = \left( e^{\beta\omega} - 1 \right)^{-1}$ is the Bose-Einstein distribution at finite temperature $T = \beta^{-1}$. 
A similar analysis applied to the dilepton production rate $\gamma \rightarrow \ell + \bar{\ell}$, leads to the formula~\cite{PhysRevD.110.056003,PhysRevD.109.056008}
\bea
\frac{d^4R_{\ell\bar{\ell}}}{dp^4}=\frac{\alpha_\text{em}}{12\pi^4}\frac{n_{\text{B}}(\omega)}{M^2}\text{Im}\left\{g_{\mu\nu}\Pi_{\text{R}}^{\mu\nu}(p)\right\},
\label{eq.Rll}
\eea
with $\alpha_\text{em}$ the QED fine structure constant, and $M$ the lepton mass.
Several previous theoretical studies have computed and analyzed the anisotropic flow coefficients in this context~\cite{PhysRevD.102.076010,PhysRevD.109.056008,Bandyopadhyay_PhysRevD.94.114034}, including scenarios involving a superposition of multiple Landau levels under the assumption of a constant magnetic field~\cite{PhysRevD.102.076010,PhysRevD.109.056008}. These works also examine the emission rates for transitions between specific pairs of levels, highlighting that the quantization effects resulting from their discrete structure are particularly significant for photon emission at low momenta. ($p_T \lesssim \sqrt{|eB|}$). In contrast, for large momenta ($p_T \gg \sqrt{|eB|}$) the anisotropy due to the effect of the magnetic field remains but resembles the classical synchrotron radiation pattern with a maximum near $\phi \sim \pi/2$~\cite{PhysRevD.102.076010,PhysRevD.109.056008,Bandyopadhyay_PhysRevD.94.114034}.

Those previous studies satisfactorily explain, at least partially, the anisotropic effects owing to the presence of a uniform background magnetic field. However, based on our previous theoretical predictions for the effects of the magnetic-field fluctuations over the photon polarization tensor~\cite{PhysRevD.109.056007}, according to Eq.~\eqref{eq:Rgamma} and Eq.~\eqref{eq.Rll} we expect that those stochastic fluctuations should also modify the corresponding predictions for the experimental photon production, particularly through the anisotropic flow coefficients. Therefore, the present work explores such a possibility, by incorporating the effects of magnetic noise within the lowest Landau level (LLL) approximation for the fermion propagators in the regime of very intense magnetic fields. In what follows, we present our theoretical results and show the predicted magnitude of those effects.

\section{Theory}
Our analysis for the noise-dependent emission rate distribution will be based on our previous results obtained for the Fermion propagator~\cite{PhysRevD.107.096014} 
\bea
    \ii S_{\Delta}(x,x')=\Phi(x,x')\int\frac{d^4p}{(2\pi)^4}e^{-\ii p\cdot(x-x')}\ii S_{\Delta}(p).
\eea
Here, the Schwinger phase factor $\Phi(x,x')$ takes the form:
\bea
 \Phi(x,x')=\exp\Bigg\{\ii e\int_x^{x'}d\xi^\mu\left[A_\mu+\frac{1}{2}F_{\mu\nu}(\xi-x')^\nu\right]\Bigg\}.
\eea
\begin{figure}[h!]
    \centering
    \includegraphics[width=0.65\linewidth]{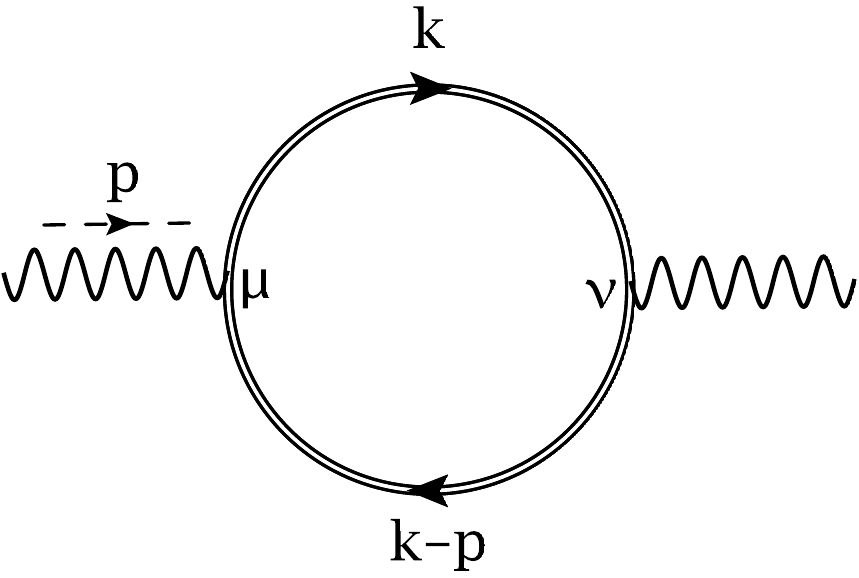}
    \caption{Feynman diagram for the polarization tensor $\Pi^{\mu\nu}(p)$ at one loop. The double lines represent the magnetic-noise-dressed Fermion propagators $S_{\Delta}(k)$ defined in Eq.~\ref{eq:SDelta}.}
    \label{fig:Polarization}
\end{figure}

For an average background magnetic field $\mathbf{B}$ oriented along the $\hat{z}$-direction, in the symmetric gauge
    \bea
    A_\mu^\text{BG}=\frac{B}{2}(0,-x_2,x_1,0),
    \eea
we explicitly obtain the Schwinger phase as
    \bea
    \Phi(x,x')=\exp\left(\frac{\ii q_f B}{2}\epsilon_{ij}x_ix_j'\right),
    \eea
where $\epsilon_{ij}$ is the two-dimensional Levy-Civita tensor.

By inspecting the magnetic field intensities in Fig.~\ref{fig:DataPhysRevC107_034901_2023}, we see that even for the lowest values $\qB/m_{\pi}^2 \simeq 10^{-2}$, we have for the light quark flavors
\bea
&&\frac{\qB}{m_u^2}\sim10^{-2}\left(\frac{m_\pi^2}{m_u^2}\right)\approx37\nn\\
&&\frac{\qB}{m_d^2}\sim10^{-2}\left(\frac{m_\pi^2}{m_d^2}\right)\approx8,
\eea
and hence we can safely assume that the phenomenological scenario of interest is compatible with the very intense magnetic field regime $|q_f B|/m_f^2 \gg 1$, thus supporting the lowest Landau Level (LLL) approximation.
On the other hand, the translational invariant factor in the propagator, for a Fermion with charge $q_f$ and mass $m_f$, can be expressed up to first order in the noise parameter $\Delta$ as~\cite{PhysRevD.109.056007,PhysRevD.107.096014}
\bea
&&\ii S_\Delta(p)=\ii S_0(p)+\ii\Delta\left(\frac{\qB}{2\pi}\right)\Big[\Theta_1(p)(\slashed{p}_\parallel+m_f)\mathcal{O}^{(+)}\nn\\
&-&\Theta_2(p)\gamma^3\mathcal{O}^{(+)}-\Theta_3(p)\sign\ii\gamma^1\gamma^2(\slashed{p}_\parallel+m_f)\Big]\nn\\
&+&\text{O}(\Delta^2),
\label{eq:SDelta}
\eea
where
\bea
\ii S_0(p)&=&2\ii\frac{e^{-\pt^2/|q_f B|}}{\pp^2-m_f^2}(\slashed{p}_\parallel+m_f)\mathcal{O}^{(+)}
\eea
is the fermion propagator in the presence of an intense magnetic field (i.e. LLL approximation). The spin-projection operator is given by
\bea
\mathcal{O}^{(\pm)}&=&\frac{1}{2}\left[1\pm\sgn(q_f B)\ii\gamma^1\gamma^2\right],
\eea
and we defined the functions~\cite{PhysRevD.109.056007,PhysRevD.107.096014}:
\begin{subequations}
\bea
\Theta_1(p)&\equiv&\frac{3(\pp^2+m_f^2)e^{-2\pt^2/|q_f B|}}{(\pp^2-m_f^2)^2\sqrt{p_0^2-m_f^2}} ,
\eea
\bea
\Theta_2(p)&\equiv&\frac{3p_3e^{-2\pt^2/|q_f B|}}{(\pp^2-m_f^2)\sqrt{p_0^2-m_f^2}},
\eea
\bea
\Theta_3(p)&\equiv&\frac{e^{-2\pt^2/|q_f B|}}{(\pp^2-m_f^2)\sqrt{p_0^2-m_f^2}}.
\eea
\end{subequations}

We shall perform our calculations at finite temperature, and hence the momentum component $p^{0} \equiv \ii\nu_{n}$, with $\nu_n = 2n\pi/\beta$ ($n\in\mathbb{Z}$) a Bosonic Matsubara frequency. The polarization tensor is then obtained by calculating the Feynman diagram depicted in Fig.~\ref{fig:Polarization} in Matsubara space $k^{\mu} = (\ii\omega_n,\mathbf{k})$
{\small
\bea
\Pi^{\mu\nu}(p)&=&-\frac{1}{2\beta}\sum_{\omega_n}\int\frac{d^3k}{(2\pi)^3}\text{Tr}\Big\{\ii q_f\gn\ii S_\Delta\left(k\right)\ii q_f\gm\ii S_\Delta(k-p)\Big\}
\label{eq:PloTenDef}.\nn\\
&+&\text{C.C.}
\eea}

The retarded polarization tensor is finally obtained by analytic continuation from the previous expression, according to the usual prescription
\be
\Pi_R^{\mu\nu}(p^0=\omega,\mathbf{p}) = \Pi^{\mu\nu}(\ii\nu_n\rightarrow \omega + \ii\epsilon,\mathbf{p}).
\ee

\section{Results}

\begin{figure*}
    \centering
    \includegraphics[scale=0.55]{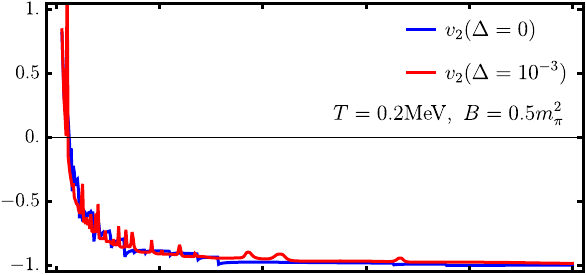}~~\includegraphics[scale=0.55]{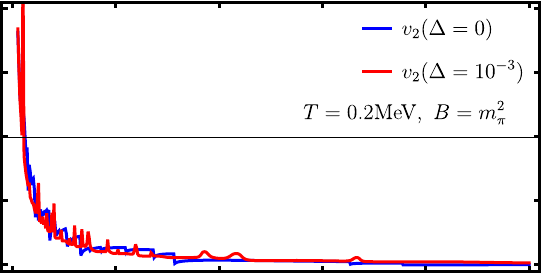}~~\includegraphics[scale=0.55]{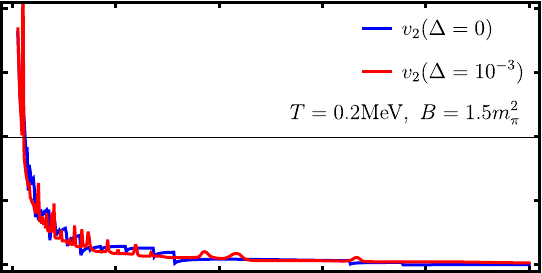}\\
    \vspace{0.1cm}
    \includegraphics[scale=0.55]{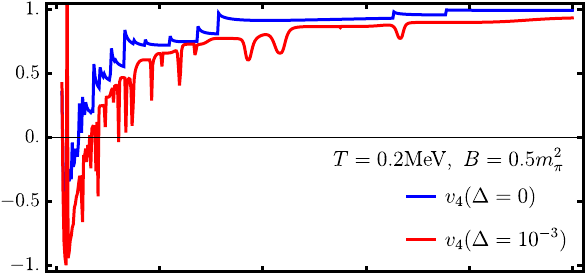}~~\includegraphics[scale=0.55]{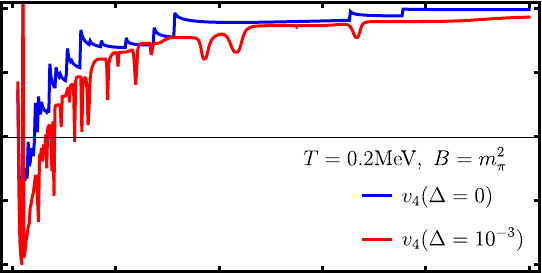}~~\includegraphics[scale=0.55]{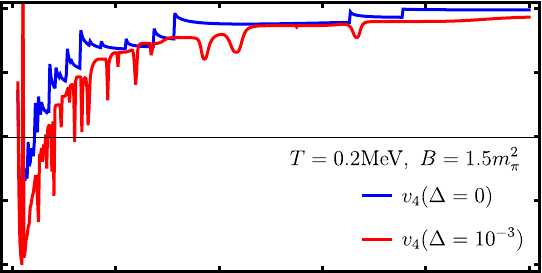}\\
    \vspace{0.1cm}
    \includegraphics[scale=0.55]{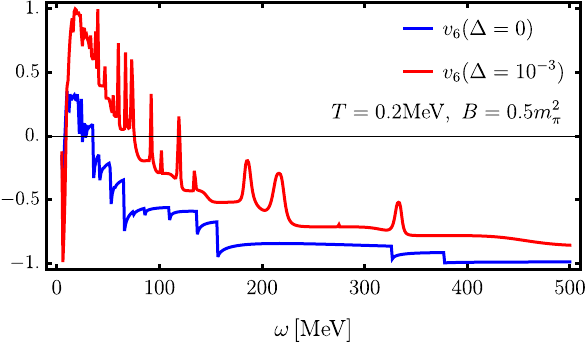}~~\includegraphics[scale=0.55]{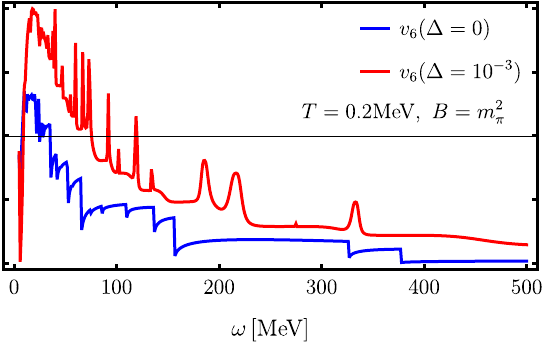}~~\includegraphics[scale=0.55]{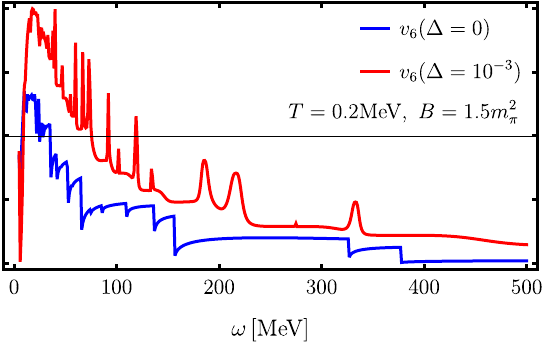}
    \caption{Anisotropic flow coefficients for emitted photons $v_2$ (top panel), $v_4$ (middle panel), and $v_6$ (bottom panel) as functions of the photon energy $\omega$, calculated from Eq.~\eqref{eq:vn} for $T = 0.2$ MeV and varying strengths of the background magnetic field. The blue line represents the noiseless limit $\Delta = 0$, while the red line corresponds to $\Delta = 10^{-3} \, \text{MeV}^{-1}$.}
    \label{fig:Flows}
\end{figure*}

\begin{figure*}
    \centering
    \includegraphics[scale=0.7]{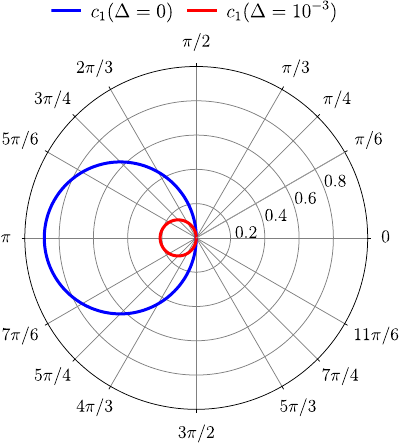}\hspace{0.8cm}\includegraphics[scale=0.7]{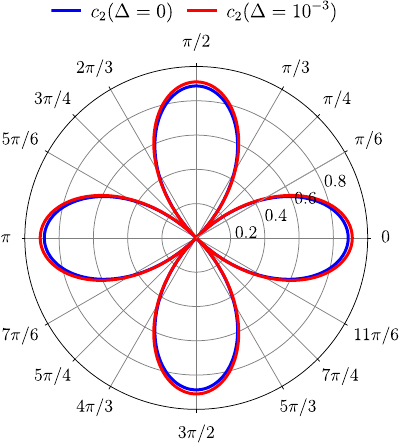}\hspace{0.8cm}\includegraphics[scale=0.7]{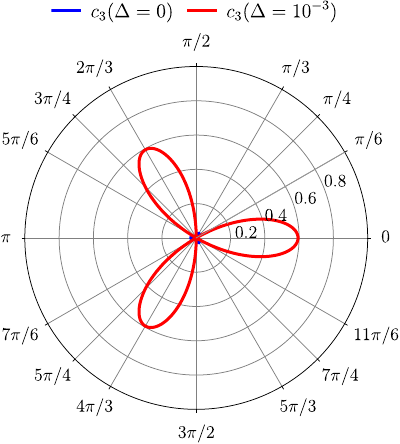}\\
    \vspace{0.5cm}
    \includegraphics[scale=0.7]{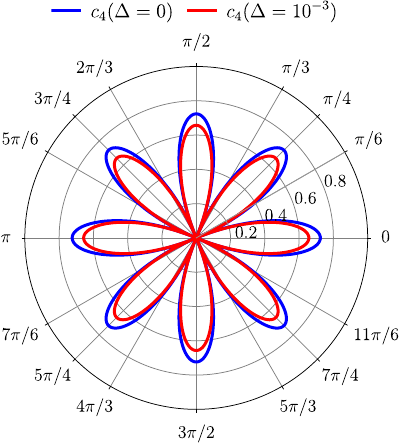}\hspace{0.8cm}\includegraphics[scale=0.7]{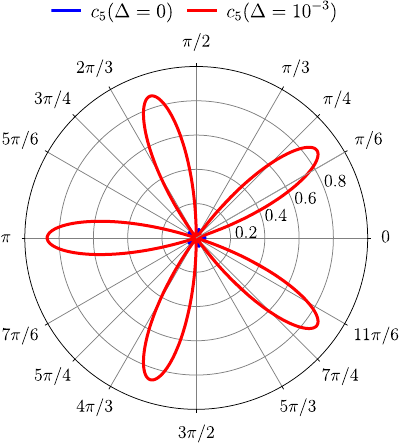}\hspace{0.8cm}\includegraphics[scale=0.7]{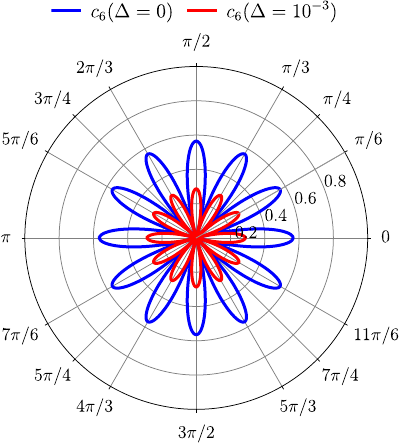}
    \caption{Angular distribution components $c_n(\phi) = v_n\cos(n\phi)$ (in polar coordinates) for emitted photons from Eq.~\eqref{ec:defflujos}, with $v_n$ calculated from Eq.~\eqref{eq:vn} for $\omega = 100$ MeV, $T = 0.2$ MeV, and $B = 0.5 m_\pi^2$. The blue line corresponds to the noiseless limit $\Delta = 0$, while the red line represents $\Delta = 10^{-3}\,\text{MeV}^{-1}$}
    \label{fig:cn}
\end{figure*}

\begin{figure*}
    \centering
    \includegraphics[scale=0.74]{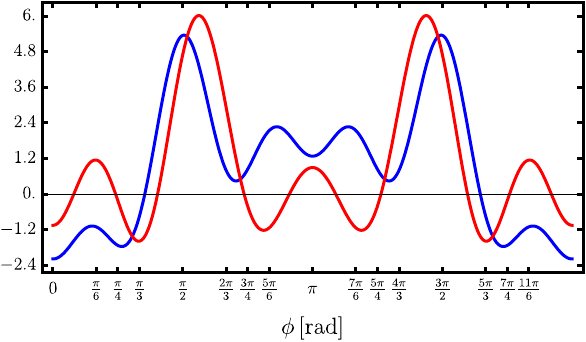}\hspace{1.5cm}\includegraphics[scale=0.74]{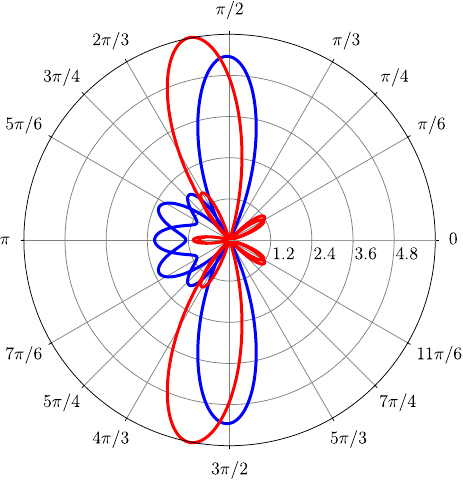}
    \caption{Total angular distribution for emitted photons from Eq.~\eqref{ec:defflujos} (normalized by $\mathcal{R}_0/2\pi$), with $v_n$ (up to $n = 6$) calculated from Eq.~\eqref{eq:vn} for $\omega = 100$ MeV, $T = 0.2$ MeV, and $B = 0.5 m_\pi^2$. In normal (left) and polar coordinates (right). The blue line corresponds to the noiseless limit $\Delta = 0$, while the red line represents $\Delta = 10^{-3}\,\text{MeV}^{-1}$.}
    \label{fig:ctotal}
\end{figure*}

\begin{figure*}
    \centering
    \includegraphics[scale=0.74]{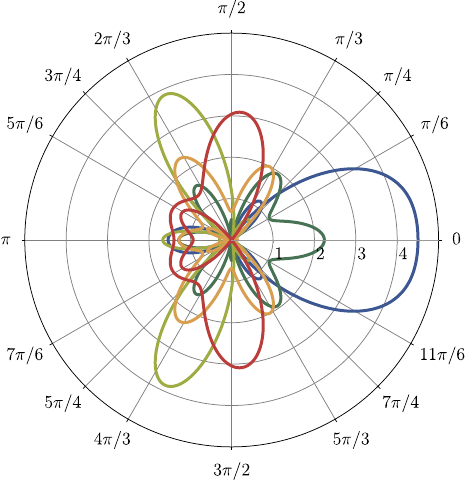}
    \hspace{1.5cm}
    \includegraphics[scale=0.74]{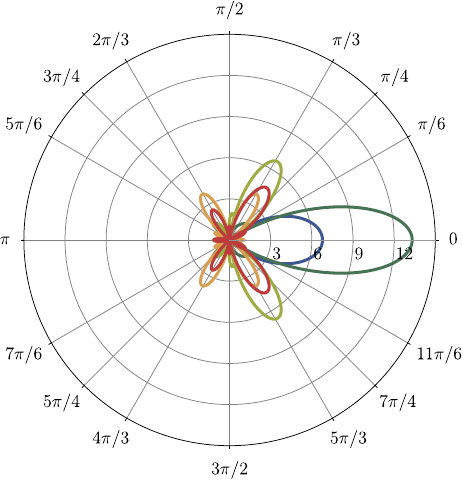}\\
    \vspace{0.3cm}
    \includegraphics[scale=0.75]{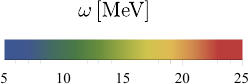}
    \caption{Total angular distribution for emitted photons (in polar coordinates) from Eq.~\eqref{ec:defflujos} (normalized by $\mathcal{R}_0/2\pi$), with $v_n$ (up to $n = 6$) calculated for $5$MeV$\leq\omega\leq25$MeV, $T = 0.2$ MeV, $B = 0.5 m_\pi^2$, in polar coordinates. The left panel represents the noiseless limit $\Delta=0$, and the right panel corresponds to $\Delta=10^{-3}$MeV$^{-1}$. Note the differences in amplitude on each plot.}
    \label{fig:ctotal_lowenergy}
\end{figure*}


\begin{figure*}
    \centering
    \includegraphics[scale=0.55]{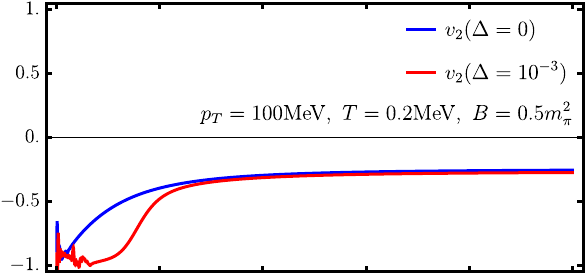}~~\includegraphics[scale=0.55]{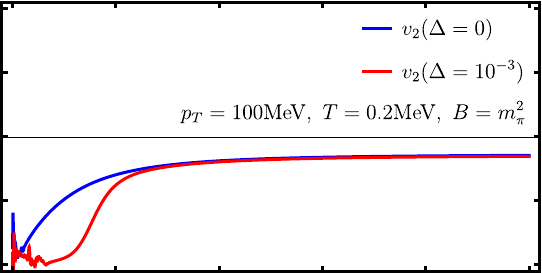}~~\includegraphics[scale=0.55]{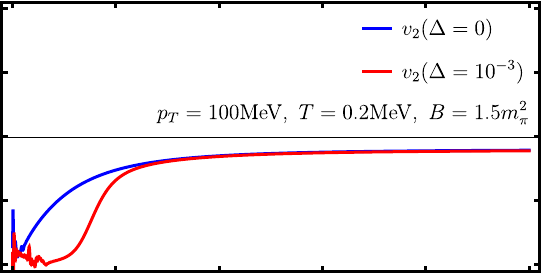}\\
    \vspace{0.1cm}
    \includegraphics[scale=0.55]{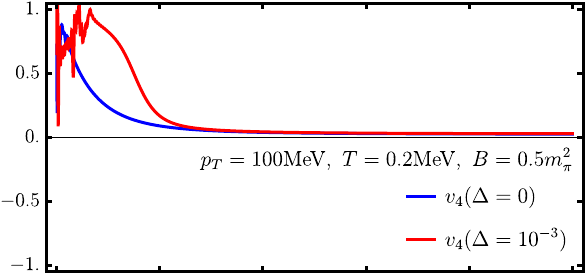}~~\includegraphics[scale=0.55]{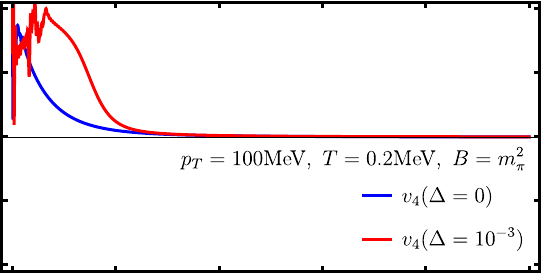}~~\includegraphics[scale=0.55]{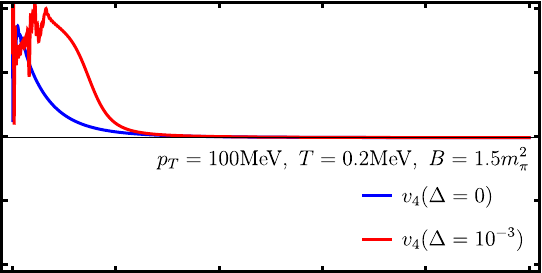}\\
    \vspace{0.1cm}
    \includegraphics[scale=0.55]{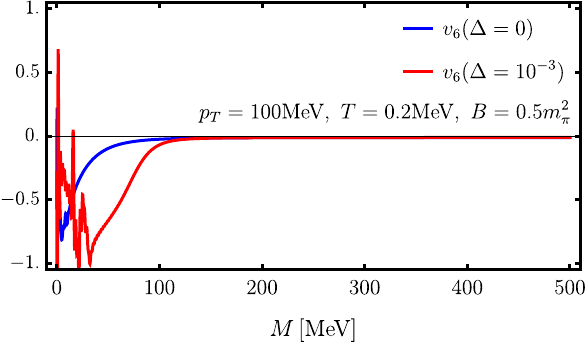}~~\includegraphics[scale=0.55]{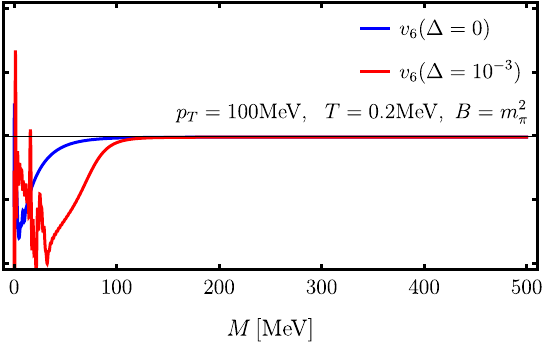}~~\includegraphics[scale=0.55]{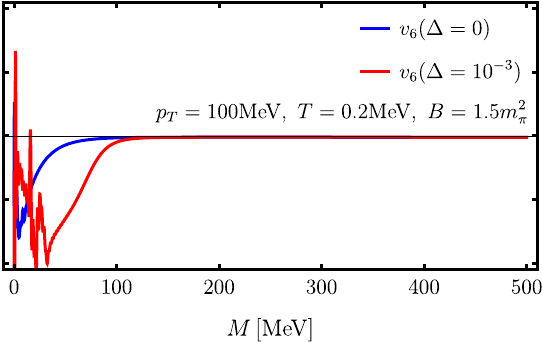}
    \caption{Dilepton anisotropic flow coefficients $v_2$ (top panel), $v_4$ (middle panel), and $v_6$ (bottom panel) as functions of the dilepton invariant mass $M$, calculated from Eq.~\eqref{eq:vn} for $T = 0.2$ MeV and varying strengths of the background magnetic field. The blue line represents the noiseless limit $\Delta = 0$, while the red line corresponds to $\Delta = 10^{-3} \, \text{MeV}^{-1}$.}
    \label{fig:Flows_dilepton}
\end{figure*}

\begin{figure*}
    \centering
    \includegraphics[scale=0.6]{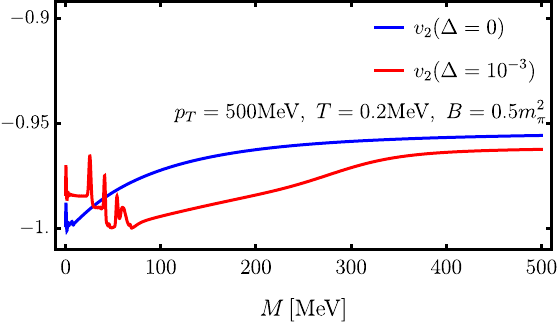}
    \hspace{0.34cm}\includegraphics[scale=0.6]{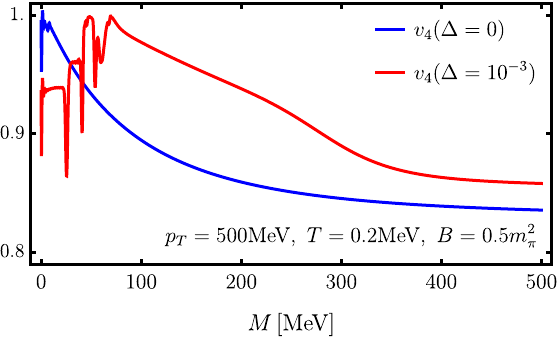}
    \hspace{0.34cm}\includegraphics[scale=0.6]{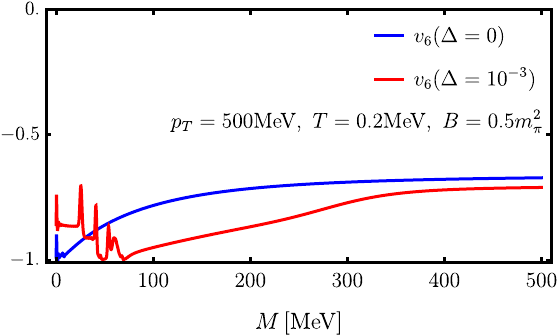}
    \caption{Dilepton anisotropic flow coefficients $v_2$ (top), $v_4$ (center), and $v_6$ (bottom) as functions of the dilepton invariant mass $M$, calculated calculated from Eq.~\eqref{eq:vn} for $T = 0.2$ MeV, $p_T = 500\,\text{MeV}$ and varying strengths of the background magnetic field. The blue line represents the noiseless limit $\Delta = 0$, while the red line corresponds to $\Delta = 10^{-3} \, \text{MeV}^{-1}$.}
    \label{fig:Flows_dilepton_highenergy}
\end{figure*}


\begin{figure*}
    \centering
    \includegraphics[scale=0.7]{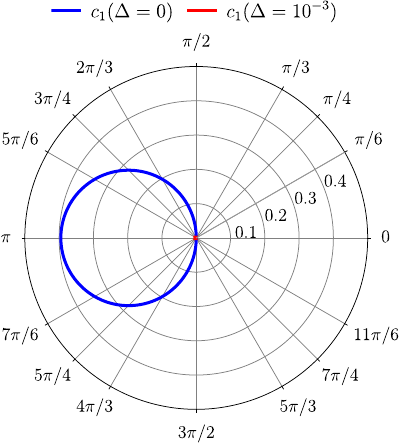}\hspace{0.8cm}\includegraphics[scale=0.7]{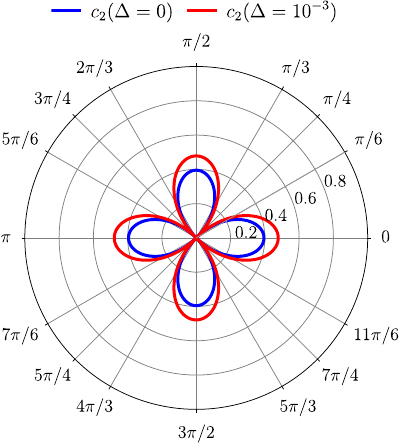}\hspace{0.8cm}\includegraphics[scale=0.7]{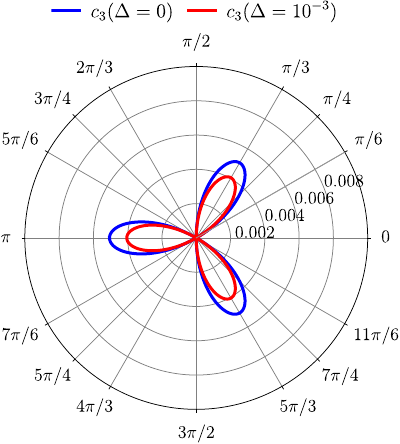}\\
    \vspace{0.5cm}
    \includegraphics[scale=0.7]{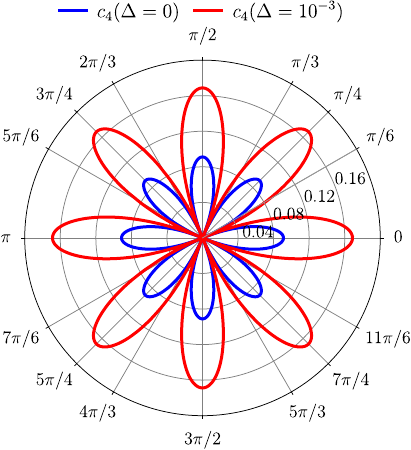}\hspace{0.8cm}\includegraphics[scale=0.7]{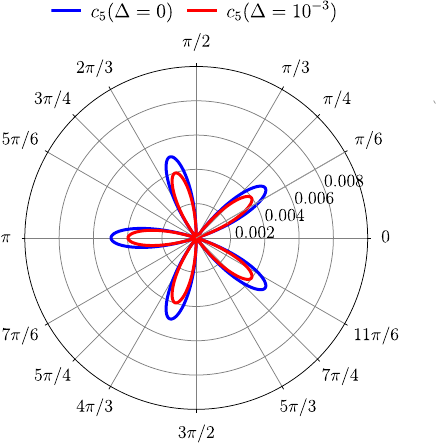}\hspace{0.8cm}\includegraphics[scale=0.7]{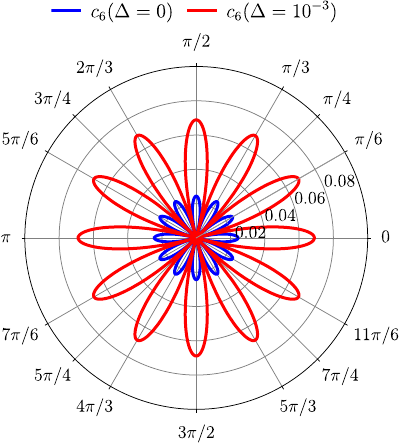}
    \caption{Angular distribution components $c_n(\phi) = v_n\cos(n\phi)$ from Eq.~\eqref{ec:defflujos} for dilepton production, with $v_n$ calculated from Eq.~\eqref{eq:vn} for $p_T=M= 100$ MeV, $T = 0.2$ MeV, and $B = 0.5 m_\pi^2$. The blue line corresponds to the noiseless limit $\Delta = 0$, while the red line represents $\Delta = 10^{-3}\,\text{MeV}^{-1}$}
    \label{fig:cn_dilepton}
\end{figure*}

\begin{figure}
    \centering
    \includegraphics[scale=0.74]{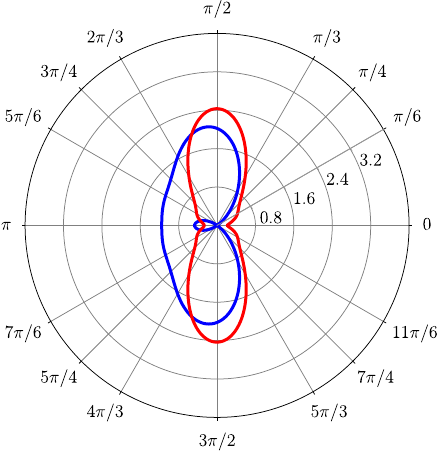}\\
    \vspace{0.3cm}\includegraphics[scale=0.74]{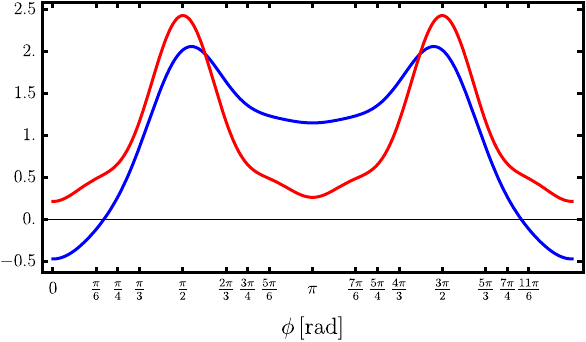}
    \caption{Dilepton total angular distribution from Eq.~\eqref{ec:defflujos} (normalized by $\mathcal{R}_0/2\pi$), with $v_n$ (up to $n = 6$) calculated for $p_T=M= 100$ MeV, $T = 0.2$ MeV, and $B = 0.5 m_\pi^2$. The blue line corresponds to $\Delta = 0$, while the red line represents $\Delta = 10^{-3}\,\text{MeV}^{-1}$.}
    \label{fig:ctotal_dilepton}
\end{figure}

\begin{figure}
    \centering
    \includegraphics[scale=0.74]{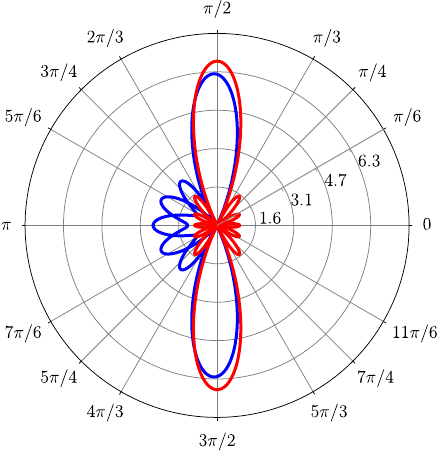}\\
    \vspace{0.3cm}\includegraphics[scale=0.74]{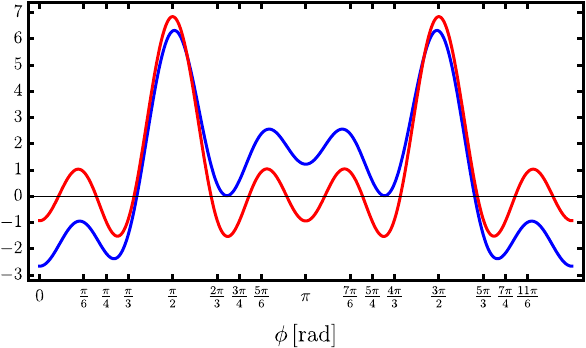}
    \caption{Dilepton total angular distribution from Eq.~\eqref{ec:defflujos} (normalized by $\mathcal{R}_0/2\pi$), with $v_n$ (up to $n = 6$) calculated for $p_T=500$MeV, $M= 100$ MeV, $T = 0.2$ MeV, and $B = 0.5 m_\pi^2$. The blue line corresponds to $\Delta = 0$, while the red line represents $\Delta = 10^{-3}\,\text{MeV}^{-1}$.}
    \label{fig:ctotal_dilepton_pT500}
\end{figure}

In this section, we define the emission rate distributions for photon and dilepton production, respectively, in the presence of stochastic magnetic fluctuations as follows
\bea
\frac{d^3R^{(\text{noise})}_{\gamma}}{p_\perp dp_\perp d\phi dy} &\equiv& \frac{d^3R^{0}_{\gamma}}{p_\perp dp_\perp d\phi dy} + \frac{d^3R^{\Delta}_{\gamma}}{p_\perp dp_\perp d\phi dy},\nn\\
\frac{d^4R_{\ell\bar{\ell}}^{(\text{noise})}}{dp^4} &\equiv& \frac{d^4R_{\ell\bar{\ell}}^{0}}{dp^4} + \frac{d^4R_{\ell\bar{\ell}}^{\Delta}}{dp^4}.
\label{eq:photdileprates}
\eea

Even though the analytical expressions obtained for both processes share many similarities, the kinematical conditions due to the finite lepton mass $M$ differ in both cases, and hence in what follows we shall analyze them separately.

\subsection{Photon emission}
The noiseless emission rate distribution for photons is given by the expression (see the Appendix~\ref{appA} for details)
\bea
\frac{d^3R^{0}_{\gamma}}{p_\perp dp_\perp d\phi dy}&=&\sum_{f=u,d}\frac{q_f^2m_f^2\qB N_c n_B(\omega)}{32\pi^4}e^{-\frac{\pt^2}{2\qB}}\nn\\
&\times&\theta\left(\omega -\sqrt{p_z^2 + 4 m_f^2} \right)\mathcal{I}_0,
\label{eq:photon_rate0}
\eea
with $\theta(x)$ the Heaviside step function,  $n_F(z) = \left( e^{\beta z} + 1  \right)^{-1}$ is the Fermi-Dirac distribution, and we defined
\bea
\mathcal{I}_0 &=& -\sum_{s_1=\pm}\sum_{s_2=\pm}\sum_{s=\pm}\theta\left(s_1 E_+^{(s)}\right)\theta\left(s_2 E_{-}^{(s)}\right)\nn\\
&&\times\frac{\left[n_\text{F}( E^{(s)}_-)-n_\text{F}( E^{(s)}_+)\right]}{E^{(s)}_+ E^{(s)}_-\left|\frac{k_s}{E^{(s)}_+}-\frac{(k_s-p)}{E^{(s)}_-}\right|}.
\label{eq:I0}
\eea

In the expression above, we also define (for $s=\pm$)
\bea
k_s=\frac{p_z}{2}+s\frac{|\omega|}{2}\sqrt{1-\frac{4 m_f^2}{\omega^2 - p_z^2}},
\label{roots1}
\eea
and
\bea
E_{\pm}^{(s)} &=&  \pm\frac{\omega}{2} + s \frac{p_z}{2} \sqrt{1 - \frac{4 m_f^2}{\omega^2 - p_z^2}}.
\eea

In addition, for the
noise-related excess yield for photon and dilepton production we obtained the analytical expression
\bea
&&\frac{d^3R^{\Delta}_{\gamma}}{p_\perp dp_\perp d\phi dy}=\frac{n_\text{B}(\omega)\Delta N_c}{32 \pi^5}\sum_{f=u,d} q_f^2\qB^2m_f^2e^{-\frac{2 \mathbf{p}_{\perp}^2}{3\qB}}\nn\\
&&\times\theta\left( \omega - \sqrt{p_z^2 + 4m_f^2} \right) \left\{ 
\mathcal{I} + \frac{3}{m_f^2}\mathcal{J}
\right\}.
\eea

\begin{widetext}
Here, we define (see the Appendix~\ref{AppB} for details)
\bea
&&\mathcal{I}\equiv -\sum_{s_1=\pm}\sum_{s_2=\pm}\sum_{s=\pm}\frac{\theta\left(s_1 E_+^{(s)}\right)\theta\left(s_2 E_{-}^{(s)}\right)\left(\left[E_{-}^{(s)}\right]^4-3 m_f^2 \left[E_{-}^{(s)}\right]^2 + m_f^4\right) \left\{n_F\left( E_{-}^{(s)} \right) - n_F\left( E_+^{(s)} \right) \right\}}{E_+^{(s)} \left[E_{-}^{(s)}\right]^3\left( \left[E_{-}^{(s)}\right]^2 - m_f^2 \right)^{3/2}\left|\frac{ k_s}{E_+^{(s)}}-\frac{(k_s-p_z)}{E_{-}^{(s)}} \right|}\nn\\
&&\mathcal{J} \equiv \sum_{s_1=\pm}\sum_{s_2=\pm}\sum_{s=\pm}\frac{\theta\left(s_1 E_+^{(s)}\right)\theta\left(s_2 E_{-}^{(s)}\right)\left[E_+^{(s)}E_{-}^{(s)}-k_s(k_s-p_z)\right]\left\{n_\text{F}\left( E_-^{(s)}\right)-n_\text{F}\left( E_+^{(s)}\right)\right\}}{E_+^{(s)}E_-^{(s)}\sqrt{\left[E_{-}^{(s)}\right]^2 - m_f^2}\left|\frac{   k_s}{E_+^{(s)}}-\frac{ (k_s-p_z)}{E_-^{(s)}} \right|},
\eea
\label{eq:IJ}
\end{widetext}

The angular distribution for the total yield is symmetric with respect to $\phi = \pi$. Moreover, in the fundamental interval $\phi \in \left[0,\pi \right]$, it exhibits three sharp singularities. The first, which is also present in the absence of noise $\Delta=0$, arises at $\phi = \pi/2$, which is the kinematical condition for $p_z = \omega$. In addition, the noise terms introduce two singularities at $\phi = \pi/2 \pm \arcsin\left(\sqrt{2 m_f /\omega} \right)$. These additional resonances correspond to the threshold $\left[E_{-}^{(s)}\right]^2 = m_f^2$.

The anisotropic flow coefficients $v_n$ computed from Eq.~\eqref{eq:vn}, for $n=2,\,4,\,6$, are displayed in the panel Fig.~\ref{fig:Flows}. The blue line represents the noiseless limit $\Delta = 0$, while the red line incorporates the effect of stochastic magnetic noise $\Delta = 10^{-3}\,\text{MeV}^{-1}$. The coefficients $v_3 = v_5 = 0$ exactly vanish, regardless of the presence of magnetic noise. Although the elliptic flow coefficient $v_2$ is very weakly sensitive to stochastic noise, the higher-order anisotropic flow coefficients $v_4$ and $v_6$ display a strong effect, particularly at low photon frequencies. 
In the panel Fig.~\ref{fig:cn}, we represent the azimuthal Fourier components $c_{n}(\phi) = v_n\cos(n\phi)$ for the distribution of the emitted radiation, according to the definition in Eq.~\eqref{ec:defflujos}. As clearly observed in the plots, the characteristic distribution of lobules for each mode is preserved in the presence of noise; however, the corresponding amplitude for each mode is affected differently. In particular, the mode $c_2(\phi)$ that characterizes the ellipticity of the flow is very weakly affected by the noise. In contrast, the higher-order modes $c_4(\phi)$ and $c_6(\phi)$ are strongly modified, even for a moderately small noise $\Delta = 10^{-3}\,\text{MeV}^{-1}$, as clearly seen in the second and third subfigures in the panel Fig.~\ref{fig:cn}. In Fig.~\ref{fig:ctotal}, we represent the total angular distribution (normalized by $\mathcal{R}_0/2\pi$) in normal (left) and polar (right) coordinates, including the superposition of the components $c_n(\phi)$ up to $n=6$, for $\omega = 100\,\text{MeV}$. The blue line represents the noiseless limit $\Delta = 0$, while the red line corresponds to a magnetic stochastic noise of magnitude $\Delta = 10^{-3}\,\text{MeV}^{-1}$.
As expected, the angular distribution displays the two large and opposite lobules near $\phi\sim\pi/2$ and $\phi\sim 3\pi/2$, respectively, which are reminiscent of the classical synchrotron radiation distribution. Remarkably, while in the noiseless limit $\Delta=0$ the main lobules are exactly aligned at these angles, in the presence of noise the direction of the lobules is slightly deflected towards the $\phi = \pi$ axis. Moreover, in these new directions the size of the lobules is larger, thus showing that noise also enhances the relative magnitude of the emitted radiation along those dominant directions. A different anisotropic radiation pattern is observed at lower photon frequencies $5\,\text{MeV}\le \omega \le 25\,\text{MeV}$, as depicted in Fig.~\ref{fig:ctotal_lowenergy} for $\Delta = 0$ (left) and $\Delta = 10^{-3}\,\text{MeV}$, respectively. In this low-energy regime,  for $\omega \lesssim 10\,\text{MeV}$ the angular distribution is first dominated by the parallel direction, as seen by a dominant lobule along $\phi = 0$ which is present both in $\Delta = 0$ and in $\Delta = 10^{-3}\,\text{MeV}^{-1}$. As the photon energy increases, the distribution gradually acquires a dominant pattern along $\phi = \pi/2$ and $\phi = 3\pi/2$, thus resembling the classical distribution already discussed in Fig.~\ref{fig:ctotal}. Remarkably, the noise effects are clearly stronger at low photon frequencies, as can be appreciated by comparing the left and right Figs.~\ref{fig:ctotal_lowenergy}, where for $\Delta = 10^{-3}\,\text{MeV}^{-1}$ the rate distribution is much more focused towards the parallel direction $\phi = 0$ than for the noiseless limit $\Delta = 0$.   

\subsection{Dilepton emission}
The magnetic stochastic noise contribution to the total dilepton production rate, as defined in Eq.~\eqref{eq:photdileprates}, is given by the expression
\bea
\frac{d^4R_{\ell\bar{\ell}}^{\Delta}}{dp^4}&\equiv&\frac{\alpha_\text{em} n_\text{B}(\omega)\Delta N_c}{48\pi^5}\sum_fq_f^2\qB\text{Im}\left\{g_{\mu\nu}\sum_{i=1}^3\tau_i^{\mu\nu}\right\},\nn\\
\label{eq:RDelta}
\eea
where the trace is over the same tensor structures (see Appendix~\ref{AppB} for details) involved in the photon emission calculation already discussed.
As mentioned before, the finite invariant lepton mass $M$ imposes further kinematic restrictions to the emission rate distribution. Nevertheless, it is possible to exploit the mathematical similarities between both photon and dilepton production rates by modifying the energy dispersion relation accordingly~\cite{PhysRevD.109.056008}
\bea
\omega\to\sqrt{M^2+p_T^2},
\eea
where $\omega$ is the photon's energy.
By following this transformation, we obtained the corresponding azimuthal distribution for dilepton production rate Eq.~\eqref{eq.Rll} from the former expression for photons, and then computed the corresponding anisotropic flow coefficients $\{ v_n \}$ as defined by Eq.~\eqref{eq:vn}. As represented in Fig.~\ref{fig:Flows_dilepton} for $p_T = 100\,\text{MeV}$ and in Fig.~\ref{fig:Flows_dilepton_highenergy} for $p_T = 500\,\text{MeV}$, respectively, as a function of the invariant lepton mass $M$, the anisotropic flow coefficients are very sensitive to the presence of magnetic noise (red line for $\Delta = 10^{-3}\,\text{MeV}^{-1}$) as compared with the noiseless limit $\Delta = 0$ (blue solid line). Remarkably, the magnitude of the average background magnetic field generates a much weaker effect over the anisotropic flow coefficients, as can be seen by comparing the different panels in Fig.~\ref{fig:Flows_dilepton}, with $B = 0.5 m_{\pi}^2$ (left), $B = m_{\pi}^2$ (center) and $B = 1.5 m_{\pi}^2$ (right). Therefore, for dilepton production it seems that stochastic magnetic noise is a major source of anisotropic effects as compared with the presence of a constant and uniform background magnetic field. The different angular components $c_n(\phi) = v_n\cos(n\phi)$, for $1\le n \le 6$ are displayed, in polar coordinates, in Fig.~\ref{fig:cn_dilepton}. From the panel, it is clear that $c_2(\phi)$, $c_3(\phi)$, and $c_5(\phi)$ exhibit a very weak noise effect. However, the remaining components, especially $c_1(\phi)$, are strongly modified by the presence of magnetic noise (red line for $\Delta = 10^{-3}\,\text{MeV}^{-1}$) as compared to the noiseless limit ($\Delta = 0$, blue line). These particular effects reveal themselves in the total emission rate, as dislplayed in Fig.~\ref{fig:ctotal_dilepton} for the condition $p_T = M = 100\,\text{MeV}$. The presence of a finite dilepton mass in the energy dispersion relation generates a more spread angular pattern (see blue line for $\Delta = 0$) as compared to the photon emission distribution. On the other hand, the presence of magnetic noise $\Delta = 10^{-3}\,\text{MeV}^{-1}$ enhances the anisotropy effects, thus creating a more focused angular pattern with lobules along $\phi = \pi/2$ and $\phi = 3\pi/2$. It is interesting to notice that, as the total momentum $p_T$ exceeds the invariant lepton mass $M$, the radiation pattern resembles more the photon case, as can be appreciated in Fig.~\ref{fig:ctotal_dilepton_pT500}, for $p_T = 500\,\text{MeV}$ and $M = 100\,\text{MeV}$, respectively. In this last figure, we observe an angular distribution remarkably similar to the photon case, where magnetic noise (red line for $\Delta = 10^{-3}$) enhances the anisotropy as compared to the noiseless limit (blue line) by focusing the radiation pattern along the characteristic lobules at $\phi = \pi/2$ and $\phi = 3\pi/2$, respectively.

\section{Discussion and Conclusions}

Motivated by the experimental evidence reported in the literature~\cite{Adare_PhysRevLett.109.122302,Adare_PhysRevC.94.064901,ALICE_2019308} for the angular anisotropy in the photon emission rate in HIC, we developed a theoretical analysis to evaluate the effects of stochastic noise in a classical magnetic field background. For this purpose, we applied our previous results for the noise-averaged fermion propagator in the limit of strong magnetic fields (i.e. LLL), in order to compute the retarded polarization tensor at finite temperature, whose imaginary part determines the angular distribution of the particle emission rates for photons and dileptons, respectively. Our analytical results thus complement previously reported studies, based on the assumption of a constant magnetic field background~\cite{PhysRevD.102.076010,PhysRevD.109.056008,Bandyopadhyay_PhysRevD.94.114034,Mizher_PhysRevD.110.L111501}, by incorporating the effects of stochastic noise as a statistical model for the nearly random anisotropies generated by the initial conditions in the nuclei participating in the individual collision events in HIC experiments. 

From our calculations, we observe that the anisotropic flow coefficients are affected by the presence of stochastic noise. Although the lobular angular pattern of the emitted radiation is preserved, qualitatively in agreement with the classical synchrotron radiation distribution, a significant deflection from the main directions $\phi = \pi/2$ and $\phi = 3\pi/2$ is observed. By analyzing the corresponding Fourier angular components, we observed that the elliptic flow coefficient $v_2$ is very weakly affected, whereas the higher-order components $v_4\cos(4\phi)$ and $v_6\cos(6\phi)$ are significantly modified by stochastic noise, thus being responsible for the anisotropy mentioned above both in the photon as well as in the dilepton production rates, respectively. From our results for the anisotropic flow coefficients, we conclude that low-frequency photons are more affected by the magnetic noise effects. This is quite intuitive, since low energy/momentum photons correspond to longer wavelength modes that are thus more susceptible to sample the spatially distributed magnetic noise. In summary, our theoretical results suggest that magnetic noise effects due to the nearly random initial conditions in the individual nuclei collision events in HIC experiments are indeed significant, and should be taken into account in the analysis of experimental signals. Despite the approximations involved in our calculations, our results imply that previously ignored effects of magnetic field noise due to nearly random initial conditions on the colliding nuclei may play a role in understanding the so-called ``photon puzzle"~\cite{PhysRevC.93.044906,SHEN2016184,PhysRevC.98.054902,Monnai_2020, PhysRevC.107.024914}.

\acknowledgements{ E.M. acknowledges financial support from Fondecyt 1230440. J.D.C.-Y. also acknowledges financial support from Fondecyt 3220087.}

%

\appendix
\onecolumngrid

\section{Contribution of the photon polarization tensor for a constant, intense magnetic field background at finite temperature and in the absense of noise $\Delta = 0$}
\label{appA}
In the lowest Landau level approximation (LLL), the photon polarization tensor reads:
\bea
\ii\Pi_0^{\mu\nu}=-\frac{1}{2}\int\frac{d^4k}{(2\pi)^4}\text{Tr}\Big\{\ii q_f\gn\ii S_0^{(-)}\left(k\right)\ii q_f\gm\ii S_0^{(-)}(k-p)\Big\}-\frac{1}{2}\int\frac{d^4k}{(2\pi)^4}\text{Tr}\Big\{\ii q_f\gn\ii S_0^{(+)}(-k+p)\ii q_f\gm\ii S_0^{(+)}(-k)\Big\},\nn\\
\label{eq:PloTenDef}
\eea
where
\bea
\ii S_0^{(\pm)}(p)&=&2\ii\frac{e^{-\pt^2/|q_f B|}}{\pp^2-m_f^2}(\slashed{p}_\parallel+m_f)\mathcal{O}^{(\pm)}
\eea
and
\bea
\mathcal{O}^{(\pm)}&=&\frac{1}{2}\left[1\pm\ii\gamma^1\gamma^2\right].
\eea

It is straightforward to show that:
\bea
g_{\mu\nu}\text{Tr}\Big\{\gn(\slashed{k}_\parallel+m_f)\Op{-}\gamma^\mu(\slashed{k}_\parallel-\slashed{p}_\parallel+m_f)\Op{-}\Big\}&=&g_{\mu\nu}\text{Tr}\Big\{\gn(-\slashed{k}_\parallel+\slashed{p}_\parallel+m_f)\Op{+}\gamma^\mu(-\slashed{k}_\parallel+m_f)\Op{+}\Big\}\nn\\
&=&4m_f^2.
\eea
Therefore, we have
\bea
g_{\mu\nu}\Pi_0^{\mu\nu}=16\ii\, q_f^2m_f^2\int\frac{d^2k_\perp}{(2\pi)^2}\exp\left[-\frac{\kt^2+(\kt-\pt)^2}{\qB}\right]\int\frac{d^2k_\parallel}{(2\pi)^2}\frac{1}{(k_\parallel^2-m_f^2)[(\kp-\pp)^2-m_f^2]}.
\eea

The perpendicular momentum integral is computed by completing the square in the exponent, and applying the Gaussian integral identity
\bea
g_{\mu\nu}\Pi_0^{\mu\nu}=\frac{8\ii\pi\, q_f^2m_f^2\qB}{(2\pi)^2}\exp\left(-\frac{\pt^2}{2\qB}\right)\int\frac{d^2k_\parallel}{(2\pi)^2}\frac{1}{(k_\parallel^2-m_f^2)[(\kp-\pp)^2-m_f^2]}.
\eea
For the parallel momenta, at finite temperature, we perform a Wick rotation to Euclidean space
\bea
k_0&\to&\ii\omega_n=2\pi\ii T(2n+1)\nn\\
p_0&\to&\ii\nu_l=2\pi\ii T l,
\label{eq:matsubarafrecuenciasdef}
\eea
so that
\bea
\int\frac{d^2k_\parallel}{(2\pi)^2}\to\ii T\sum_{n=-\infty}^{+\infty}\int\frac{dk_z}{2\pi}.
\eea

\bea
g_{\mu\nu}\Pi_0^{\mu\nu}=-\frac{8\pi\, q_f^2m_f^2\qB}{(2\pi)^2}\exp\left(-\frac{\pt^2}{2\qB}\right)T\sum_{n=-\infty}^{+\infty}\int\frac{dk_z}{2\pi}\frac{1}{[(\ii\wn)^2-k_z^2-m_f^2][(\ii\omega_n-\ii\nu_l)^2-(k_z-p_z)^2-m_f^2]},
\eea

The Matsubara sum can be performed by standard methods, to obtain the explicit result~\cite{lebellac,gluon_thermal_ayala}:
\bea
T\sum_{n=-\infty}^{+\infty}\frac{1}{[\wn+k_z^2+m_f^2][\left(\ii\omega_n-\ii\nu_l\right)^2+(k_z-p_z)^2+m_f^2]}&=&\sum_{s_1,s_2=\pm1}\frac{s_1s_2}{4E_kE_{kp}}\left[\frac{1-n_\text{F}(s_1E_k)-n_\text{F}(-s_2E_{kp})}{\ii\nu_l-s_1E_k+s_2E_{kp}}\right]\nn\\
&=& \sum_{s_1,s_2=\pm1}\frac{s_1s_2}{4E_kE_{kp}}\left[\frac{n_\text{F}(s_2E_{kp})-n_\text{F}(s_1E_k)}{\ii\nu_l-s_1E_k+s_2E_{kp}}\right].
\eea
Here, we defined the energy dispersion relations
\bea
E_k &=& \sqrt{k_z^2 + m_f^2}\nn\\
E_{kp} &=& \sqrt{(k_z - p_z)^2 + m_f^2}.
\eea
Substituting the later, we are left with a single momentum integral over $k_z$,
\bea
g_{\mu\nu}\Pi_0^{\mu\nu}=-\frac{8\pi\, q_f^2m_f^2\qB}{(2\pi)^2}\exp\left(-\frac{\pt^2}{2\qB}\right)\sum_{s_1,s_2=\pm1}\int\frac{dk_z}{2\pi}\frac{s_1s_2}{4E_kE_{kp}}\left[\frac{n_\text{F}(s_2E_{kp})-n_\text{F}(s_1E_k)}{\ii\nu_l-s_1E_k+s_2E_{kp}}\right].
\eea

Returning back to the Minkowsky space via analytic continuation $\ii\nu_l\to\omega+\ii\epsilon$ we obtain
\bea
g_{\mu\nu}\Pi_0^{\mu\nu}=-\frac{8\pi\, q_f^2m_f^2\qB}{(2\pi)^2}\exp\left(-\frac{\pt^2}{2\qB}\right)\sum_{s_1,s_2=\pm1}\int\frac{dk_z}{2\pi}\frac{s_1s_2}{4E_kE_{kp}}\left[\frac{n_\text{F}(s_2E_{kp})-n_\text{F}(s_1E_k)}{\omega-s_1E_k+s_2E_{kp}+\ii\epsilon}\right],
\eea
so that by applying the Plemelj's identity
\bea
\lim_{\epsilon\to0}\frac{1}{A\pm\ii\epsilon}=\text{PV}\left(\frac{1}{A}\right)\mp\ii\pi\delta(A),
\label{PVquotient}
\eea
the imaginary part of the above is
\bea
&&\text{Im}\left[g_{\mu\nu}\Pi_0^{\mu\nu}\right]=\frac{8\pi^2\, q_f^2m_f^2\qB}{(2\pi)^2}\exp\left(-\frac{\pt^2}{2\qB}\right)\sum_{s_1,s_2=\pm1}\int\frac{dk_z}{2\pi}\frac{s_1s_2}{4E_kE_{kp}}\left[n_\text{F}(s_2\beta E_{kp})-n_\text{F}(s_1\beta E_k)\right]\nn\\
&&\hspace{11cm}\times\delta(\omega-s_1E_k+s_2E_{kp})
\eea
The momentum integral with the delta-distribution is evaluated by means of the standard identity
\bea
\int dk_z f(k_z)\delta[h(k_z)]=\sum_i\frac{f(k_i)}{\left|h'(k_i)\right|},
\label{eq:diraccompuesta}
\eea
where $k_i$ are the roots of $h(z)$, and
\bea
\int dk_z f(k_z)\delta[h(k_z)]=0,
\eea
if $h(k_z)$ has not roots.

The general form of $h(k_z)$ is given by
\bea
h(k_z)=s_1\sqrt{k^2+m_f^2}-s_2\sqrt{(k_z-p_z)^2+m_f^2}-\omega,~\text{for}~s_1,s_2=\pm1,
\eea
so that
\bea
\frac{dh}{dk_z}=\frac{s_1k_z}{E_k}-\frac{s_2(k_z-p_z)}{E_{kp}},~\text{for}~s_1,s_2=\pm1
\eea

If the combination of signs is consistent, the roots of $h(k_z)$ are given by
\bea
k_\pm=\frac{p_z}{2}\pm\frac{|\omega|}{2}\sqrt{\frac{(\omega^2 - p^2_z-4 m_f^2)}{(\omega^2 - p_z^2)}}.
\label{roots1}
\eea
Clearly, this solution implies a threshold for pair production, such that $|\omega|> \sqrt{p_z^2 + 4 m_f^2}$.
It is more convenient, however, to evaluate the energy dispersions at this particular solution in the following way,
\bea
s_1 E_k(k_s) - s_2 E_{kp}(k_s) &=& \omega\\
s_1 E_k(k_s) + s_2 E_{kp}(k_s) &=& \Lambda_s.
\eea
We determine the value of $\Lambda_s$ by multiplying both equations, thus obtaining
\bea
(s_1 E_k)^2 - (s_2 E_{kp})^2 = \omega \Lambda_s,
\eea
from which we solve for $\Lambda_s$ (for each root $k_s$, $s=\pm$)
\bea
\Lambda_s = \frac{\left[E_k(k_s)\right]^2 - \left[E_{kp}(k_s)\right]^2}{\omega} = s p_z \sqrt{1 - \frac{4 m_f^2}{\omega^2 - p_z^2}}.
\eea
Therefore, the energy dispersion relations evaluated right at the condition defined by the support of the $\delta$-function are
\bea
s_1 E_{k}(k_s) &=& \frac{\omega}{2}+\frac{\Lambda_s}{2} = \frac{\omega}{2} + s \frac{p_z}{2} \sqrt{1 - \frac{4 m_f^2}{\omega^2 - p_z^2}}\equiv E_+^{(s)}\nn\\
s_2 E_{kp}(k_s) &=& -\frac{\omega}{2}+\frac{\Lambda_s}{2} = -\frac{\omega}{2} + s \frac{p_z}{2} \sqrt{1 - \frac{4 m_f^2}{\omega^2 - p_z^2}}\equiv E_-^{(s)}
\label{eq:E1E2}
\eea

\bea
&&\text{Im}\left[g_{\mu\nu}\Pi_0^{\mu\nu}\right]=\frac{q_f^2m_f^2\qB}{4\pi}\exp\left(-\frac{\pt^2}{2\qB}\right)\Theta\left(\omega -\sqrt{p_z^2 + 4 m_f^2} \right)\sum_{s_1,s_2=\pm1}\sum_{s=\pm}\frac{\left[n_\text{F}(\beta s_2 E_{kp})-n_\text{F}(\beta s_1 E_k)\right]}{s_1E_k s_2 E_{kp}\left|\frac{s_1k}{E_k}-\frac{s_2(k-p)}{E_{kp}}\right|}\Bigg|_{k=k_s},
\eea

where the $k_s$ are given by Eq.~\eqref{roots1}. Clearly, if $\omega-s_1E_k+s_2E_{kp}\neq0$, the integral vanishes. Finally, by recognizing that the last expression depends only on the two energy values $E_+^{(s)} = s_1E_k(k_s)$ and 
$E_-^{(s)}=s_2E_{kp}(k_s)$ as defined in Eq.~\eqref{eq:E1E2}, and recognizing the obvious constraints $s_1 E_{+}^{(s)} = E_k(k_s)>0$ and $s_2 E_{-}(k_s) = E_{kp}(k_s)>0$, the previous expression can be written in the final form quoted in the main text, Eq.~\eqref{eq:photon_rate0} and Eq.~\eqref{eq:I0}.

\section{Computation of $\ii\Pi_f^{\mu\nu}(\omega,\mathbf{p},T)$}
\label{AppB}

As we demonstrate in Ref.~\cite{PhysRevD.109.056007}, up to order $\Delta$, we obtain that the polarization tensor is given by the expression
\bea
\ii\Pi_f^{\mu\nu}&=&\ii\Pi_0^{\mu\nu}+\ii\frac{q^2_f\qB\Delta}{4\pi}\sum_{i=1}^3\tau_i^{\mu\nu},
\eea
where
    \begin{subequations}
  \bea
\tau_1^{\mu\nu}=16\ii\int\frac{d^4k}{(2\pi)^4}\frac{e^{-\kt^2/\qB}}{\kp^2-m_f^2}\Theta_1(k-p)\left[\left(\kp\cdot(\pp-\kp)+m_f^2\right)g_\parallel^{\mu\nu}+(\kp^\mu - \pp^\mu)\kp^\nu+\kp^\mu(\kp^\nu-\pp^\nu)\right]
\eea
\bea
\tau^{\mu\nu}_2=16\ii\int\frac{d^4k}{(2\pi)^4}\frac{ e^{-k_\perp^2/\qB}}{k_\parallel^2-m_f^2}\Theta_2(k-p)\left(k^3g_\parallel^{\mu\nu}+\kp^\mu \delta_3^\nu+\kp^\nu \delta_3^\mu\right)
\eea
\bea
\tau_3^{\mu\nu} = 16\ii\int\frac{d^4k}{(2\pi)^4}\frac{e^{-\kt^2/\qB}}{\kp^2-m_f^2}\Theta_3(k-p)\Bigg[(m_f^2+\kp\cdot(\pp-\kp))(g_\parallel^{\mu\nu}-g_\perp^{\mu\nu})+(\kp^\mu - \pp^\mu)\kp^\nu+ \kp^\mu(\kp^\nu - \pp^\nu)\Bigg].\nn\\
\eea
\end{subequations}

\subsection{Computation of $g_{\mu\nu}\tau_1^{\mu\nu}$}\label{sec:ComputationOfTau1}
After integrating the perpendicular momenta $k_{\perp} = (k^1,k^2)$, the tensor structure $\tau_1^{\mu\nu}$ takes the form:
\bea
\tau_1^{\mu\nu}&=&\frac{16\ii\pi\qB}{(2\pi)^2} \exp\left(-\frac{2\pt^2}{3\qB}\right)\int\frac{d^2k_\parallel}{(2\pi)^2}\frac{\left[(k-p)_\parallel^2+m_f^2\right]\left[\left(\pp\cdot\kp-\kp^2+m_f^2\right)g_\parallel^{\mu\nu}+2\kp^\mu\kp^\nu-\pp^\mu\kp^\nu-\pp^\nu\kp^\mu\right]}{(\kp^2-m_f^2)\left[(k-p)_\parallel^2-m_f^2\right]^2\sqrt{(k_0-p_0)^2-m_f^2}},\nn\\
\eea
and therefore
\bea
g_{\mu\nu}\tau_1^{\mu\nu}&=&\frac{32\ii \pi m_f^2\qB}{(2\pi)^2} \exp\left(-\frac{2\pt^2}{3\qB}\right)\int\frac{d^2k_\parallel}{(2\pi)^2}\frac{(k-p)_\parallel^2+m_f^2}{(\kp^2-m_f^2)\left[(k-p)_\parallel^2-m_f^2\right]^2\sqrt{(k_0-p_0)^2-m_f^2}}.
\eea

After performing a Wick rotation onto Euclidean space, we have
\bea
k_0&\to&\ii\omega_n=2\pi\ii T(2n+1),\,\,\,n\in\mathbb{Z}\nn\\
p_0&\to&\ii\nu_l=2\pi\ii T l,\,\,\,l\in\mathbb{Z},
\label{eq:matsubarafrecuenciasdef}
\eea
so that the integral over $k_0$ reduces to an infinite Matsubara sum as follows
\bea
\int\frac{d^2k_\parallel}{(2\pi)^2}\to\ii T\sum_{n=-\infty}^{+\infty}\int\frac{dk_z}{2\pi}.
\eea
Therefore, we have the tensor expression
{\small
\bea
g_{\mu\nu}\tau_1^{\mu\nu}=-\frac{32\pi m_f^2\qB }{(2\pi)^2}e^{-\frac{2\pt^2}{3\qB}}\int\frac{dk_z}{2\pi}T\sum_{n=-\infty}^{+\infty}\frac{1}{\sqrt{(\ii\omega_n-\ii\nu_l)^2-m_f^2}}\frac{(\ii\omega_n-\ii\nu_l)^2-(k_z-p_z)^2+m_f^2}{\left[(\ii\omega_n)^2-k_z^2-m_f^2\right]\left[(\ii\omega_n-\ii\nu_l)^2-(k_z-p_z)^2-m_f^2\right]^2},\nn\\
\eea
}
Let us define the infinite Matsubara sum
\bea
S_1=T\sum_{n=-\infty}^{+\infty}\frac{1}{\sqrt{(\ii\omega_n-\ii\nu_l)^2-m_f^2}}\frac{(\ii\omega_n-\ii\nu_l)^2-(k_z-p_z)^2+m_f^2}{\left[(\ii\omega_n)^2-k_z^2-m_f^2\right]\left[(\ii\omega_n-\ii\nu_l)^2-(k_z-p_z)^2-m_f^2\right]^2}\equiv \sum_{n=-\infty}^{+\infty} f_1(\ii\omega_n).
\label{eq:sum1}
\eea

Here, we defined the complex function
\bea
f_1(z) = T \frac{(z - \ii\nu_l)^2 - E_{kp}^2 + 2 m_f^2}{\left[ z^2 - E_k^2\right]\left[ (z - \ii\nu_l)^2 - E_{kp}^2 \right]^2\sqrt{(z - \ii\nu_l)^2 - m_f^2}}
\eea
and the dispersion relations
\bea
E_k &=& \sqrt{k_z^2+m_f^2}\nn\\
E_{kp} &=& \sqrt{(k_z-p_z)^2+m_f^2}.
\eea

In order to regularize the sum, we introduce the meromorphic function:
\bea
g(z)=-\frac{\beta}{e^{\beta z}+1},
\eea
that possesses infinitely many single poles along the imaginary
axis at the Matsubara frequencies $z_n=\ii\wn$ with residue 1. 

\begin{figure}[h!]
    \centering
    \includegraphics[scale=1]{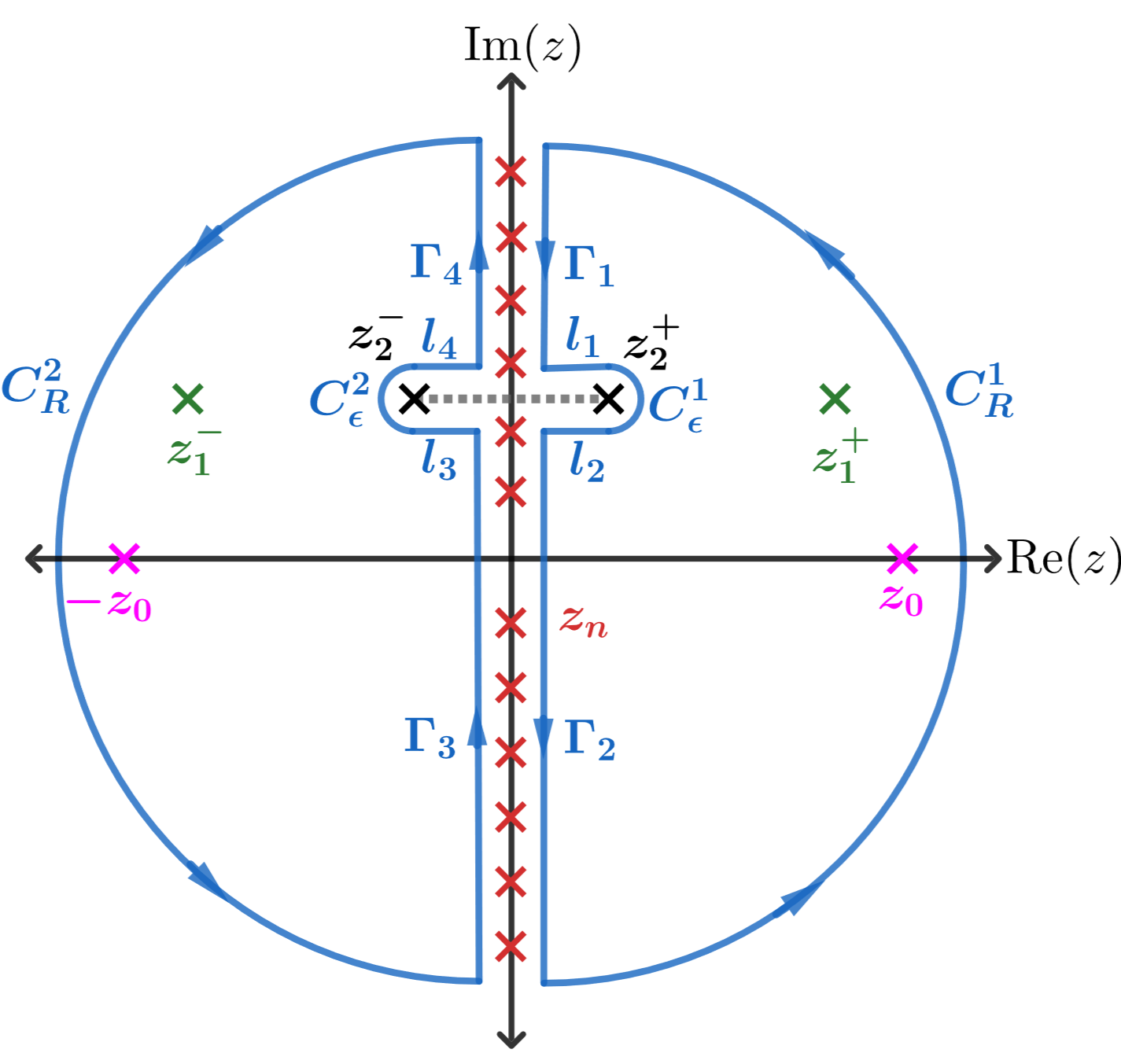}
    \caption{Contour for complex integration of Eq.~\eqref{eq:S1}. The figure depicts the poles and the branch cut of $f_1(z)$, given by the values of Eq.~\eqref{eq:poles_and_branch_1}.}
    \label{fig:contorno_suma_1}
\end{figure}

Note that, according to the Cauchy's residue Theorem
\bea
\frac{1}{2\pi\ii}\oint dz g(z)f_1(z)&=&\sum_{z_q    =z_0,z_1^{\pm},z_2^{\pm}}\left.\text{Res}\left\{g(z)f_1(z)\right\}\right|_{z_q}
\label{eq:S1}
\eea
where the integration runs in the complex plane along the contour depicted in Fig.~\ref{fig:contorno_suma_1}, and we have defined the poles
\bea
z_0&=&\sqrt{k_z^2+m_f^2} \equiv E_k \nn\\
z_1^{\pm}&=&\ii\nu_l\pm\sqrt{(k_z-p_z)^2+m_f^2} \equiv \ii\nu_l\pm E_{kp}\nn\\
z_2^{\pm}&=&\ii\nu_l\pm m_f,
\label{eq:poles_and_branch_1}
\eea
so that
\bea
\sum_{z_q    =z_0,z_1^{\pm},z_2^{\pm}}\left.\text{Res}\left\{g(z)f_1(z)\right\}\right|_{z_q}=g(z_0)\text{Res} [f_1(z)]_{z_0}+g(-z_0)\text{Res} [f_1(z)]_{-z_0}+g(z_1^+)\text{Res} [f_1(z)]_{z_1^+}+g(z_1^-)\text{Res} [f_1(z)]_{z_1^-}.\nn\\
\eea

On the other hand, in the limit $R\to\infty$
\bea
\frac{1}{2\pi\ii}\oint dz g(z)f_1(z)&=&\frac{1}{2\pi\ii}\int_\text{path 1}dz g(z)f_1(z)+\frac{1}{2\pi\ii}\int_\text{path 2}dz g(z)f_1(z),
\eea
where
\bea
\text{path 1}&=&\bigcup_{i=1}^4\Gamma_i\nn\\
\text{path 2}&=&\left(\bigcup_{i=1}^4 l_i\right)\cup\left(\bigcup_{i=1}^2C_\epsilon^i\right).
\eea

For the path 1, the Cauchy theorem implies that (reversing the orientation of the contour)
\bea
\frac{1}{2\pi\ii}\int_\text{path 1}dz g(z)f_1(z)=-\sum_n \left.\text{Res}\left\{ g(z) f_1(z)\right\}\right|_{z=\ii\omega_n} = -\sum_n f_1(\ii\omega_n)=-S_1,
\eea
provided by the fact that $f(z)$ is analytical at $z_n = \ii\omega_n$, and that $\left.\text{Res}\left\{g(z)\right\}\right|_{z = \ii\omega_n}=1$. Therefore, we have
\bea
-S_1+\frac{1}{2\pi\ii}\int_\text{path 2}dz g(z)f_1(z)=\sum_{z_q    =z_0,z_1^{\pm},z_2^{\pm}}\left.\text{Res}\left\{g(z)f_1(z)\right\}\right|_{z_q}
\eea

The path 2 has a branch cut, where the semi-circles parametrize as
\bea
C_\epsilon^1\to |z-z_2^+|=\epsilon,\text{ and }C_\epsilon^2\to |z-z_2^-|=\epsilon,
\eea
the line $l_4\cup l_1$ will be:
\bea
z=x+\ii\nu_l, \text{ with } -m_f+\epsilon\leq x\leq m_f-\epsilon,
\label{parametrization1a}
\eea
so that
\bea
\text{arg}(z-z_2^+)&=&\pi\nn\\
\text{arg}(z-z_2^-)&=&0,
\eea
and the line $l_2\cup l_3$ is:
\bea
z=x+\ii\nu_l, \text{ with } m_f-\epsilon\leq x\leq -m_f+\epsilon,
\label{parametrization1b}
\eea
so that
\bea
\text{arg}(z-z_2^+)&=&2\pi\nn\\
\text{arg}(z-z_2^-)&=&\pi.
\eea

From the latter, for $l_4\cup l_1$:
\bea
\frac{1}{\sqrt{(z-z_2^+)(z-z_2^-)}}&=&\frac{1}{\sqrt{e^{\ii\pi}|z-z_2^+|e^0|z-z_2^-|}}=\frac{1}{\ii\sqrt{|z-z_2^+||z-z_2^-|}},
\eea
and for $l_2\cup l_3$:
\bea
\frac{1}{\sqrt{(z-z_2^+)(z-z_2^-)}}&=&\frac{1}{\sqrt{e^{2\ii\pi}|z-z_2^+|e^{\ii\pi}|z-z_2^-|}}\nn\\
&=&\frac{1}{e^{3\ii\pi/2}\sqrt{|z-z_2^+||z-z_2^-|}}\nn\\
&=&-\frac{1}{\ii\sqrt{|z-z_2^+||z-z_2^-|}}
\eea
so that, in the parametrization of Eqs.~\eqref{parametrization1a} and~\eqref{parametrization1b}:
\bea
\frac{1}{\sqrt{|z-z_2^+||z-z_2^-|}}&=&\frac{1}{\sqrt{|x-m_f||x+m_f|}}=\frac{1}{\sqrt{x^2-m_f^2}}.
\eea

Hence
\bea
\frac{1}{2\pi\ii}\int_{\text{path 2}}dz g(z)f_1(z)=\frac{T}{2\pi\ii}\int_{-m_f+\epsilon}^{m_f+\epsilon}dx\frac{\left[x^2-(k_z-p_z)^2+m_f^2\right]g(x+\ii\nu_l)}{\left[(x+\ii\nu_l)^2-(k_z^2+m_f^2)\right]\left[x^2-(k_z-p_z)^2-m_f^2\right]^2}\left(\frac{-2\ii}{\sqrt{x^2-m_f^2}}\right),
\eea
so that in the limit $\epsilon\to0$:
\begin{small}
\bea
S_1=-\sum_{z_q    =z_0,z_1^{\pm},z_2^{\pm}}\left.\text{Res}\left\{g(z)f_1(z)\right\}\right|_{z_q}-\frac{T}{\pi}\text{PV}\int_{-m_f}^{m_f}dx\frac{\left[x^2-(k_z-p_z)^2+m_f^2\right]g(x)}{\left[(x+\ii\nu_l)^2-(k_z^2+m_f^2)\right]\left[x^2-(k_z-p_z)^2-m_f^2\right]^2\sqrt{x^2-m_f^2}},
\eea
\end{small}
where it has been used the fact that $g(x+\ii\nu_l)=g(x)$.
Moreover, the same property implies
\bea
g(\pm z_0) &=& g(\pm E_k) = -\beta n_\text{F}(\pm\beta E_k)\nn\\
g(z_1^{\pm}) &=& g(\ii \nu_l \pm E_{kp}) = -\beta n_\text{F}(\pm\beta E_{kp}).
\eea

Therefore, we obtain the closed expression (after analytic continuation $\ii\nu_l\rightarrow\omega+\ii\epsilon$)
\bea
\sum_{z_q    =z_0,z_1^{\pm},z_2^{\pm}}\left.\text{Res}\left\{g(z)f_1(z)\right\}\right|_{z_q}&=&-\sum_{s_1=\pm}\frac{\left[\left( \omega - s_1 E_k \right)^2 - E_{kp}^2 + 2 m_f^2\right]n_\text{F}(\beta s_1 E_k)}{2 s_1 E_k \left[E_{kp}^2-(\omega+\ii\epsilon-s_1 E_k)^2\right]^2\sqrt{\left( \omega + \ii\epsilon - s_1 E_k  \right)^2 - m_f^2}} \nn\\
&&-\sum_{s_2=\pm}\frac{\sigma_{s_2}(k_z)n_\text{F}(\beta s_2 E_{kp})}{2 s_2 E_{kp}^3\left(E_{kp}^2-m_f^2\right)^{3/2}\left[E_k^2-(\omega+\ii\epsilon+s_2 E_{kp})^2\right]^2},
\eea
where we defined
\bea
\sigma_{s_2}(k_z,\omega)&\equiv&(\omega+s_2 E_{kp})\Bigg[s_2 E_{kp}\Big((E_{kp}^2-m_f^2)^2-3m_f^2E_{kp}^2+2m_f^4\Big)+\Big((E_{kp}^2-m_f^2)^2-m_f^2E_{kp}^2\Big)\omega\Bigg]\nn\\
&-&\Big[(E_{kp}^2-m_f^2)^2-m_f^2E_{kp}^2\Big]E_k^2.
\eea

In order to extract the real and imaginary parts of the previous expressions, we shall apply the Plemelj's formula
\bea
\lim_{\epsilon\to0}\frac{1}{A\pm\ii\epsilon}=\text{PV}\left(\frac{1}{A}\right)\mp\ii\pi\delta(A),
\label{PVquotient}
\eea
and its corresponding generalization for quadratic forms
\bea
\lim_{\epsilon\to0}\frac{1}{\left(A\pm\ii\epsilon\right)^2}=-\frac{\partial}{\partial A}\lim_{\epsilon\to0}\frac{1}{A \pm \ii\epsilon}  =\text{PV}\left(\frac{1}{A^2}\right)\pm\ii\pi\delta'(A).
\label{PVquotient2}
\eea

Other forms are handled by partial fraction decomposition, in particular
\bea
\lim_{\epsilon\rightarrow 0}\frac{1}{\left(A + \ii\epsilon\right)\left(B + \ii\epsilon\right)} &=& \lim_{\epsilon\rightarrow 0}\frac{1}{B-A}\left( \frac{1}{A+\ii\epsilon} - \frac{1}{B+\ii\epsilon} \right)\nn\\
&=& PV\left( \frac{1}{AB} \right) -\ii\pi\frac{\delta(A)}{B-A} +\ii\pi\frac{\delta(B)}{B-A}
\eea
and similarly
\bea
\lim_{\epsilon\rightarrow 0}\frac{1}{\left(A + \ii\epsilon\right)^2\left(B + \ii\epsilon\right)^2} &=& \lim_{\epsilon\rightarrow 0}\frac{1}{\left(B-A\right)^2}\left( \frac{1}{A+\ii\epsilon} - \frac{1}{B+\ii\epsilon} \right)^2\nn\\
&=& \lim_{\epsilon\rightarrow 0}\frac{1}{\left(B-A\right)^2}\left( \frac{1}{\left(A+\ii\epsilon\right)^2} + \frac{1}{\left(B+\ii\epsilon\right)^2} -\frac{2}{\left(A + \ii\epsilon\right)\left(B+\ii\epsilon\right)}\right)\nn\\
&=& \frac{1}{\left( B-A \right)^2}PV\left( \frac{1}{A^2} + \frac{1}{B^2} - \frac{2}{AB}  \right)\nn\\
&+& \frac{\ii\pi}{\left( B - A \right)^2}\left[ \delta'(A) + \delta'(B) + 2\frac{\delta(A)}{B-A} - 2\frac{\delta(B)}{B-A} \right]\nn\\
&=& PV\left( \frac{1}{A^2 B^2} \right)  + \frac{\ii\pi}{\left( B - A \right)^2}\left[ \delta'(A) + \delta'(B) + 2\frac{\delta(A)}{B-A} - 2\frac{\delta(B)}{B-A} \right]
\eea
By simply factorizing the denominator, we apply the last formula as follows
\bea
&&\lim_{\epsilon\rightarrow 0}\frac{1}{\left[E_{kp}^2-(\omega+\ii\epsilon-s_1 E_k)^2\right]^2} = \lim_{\epsilon\rightarrow 0}\frac{1}{\left[ \omega + E_{kp} - s_1 E_k + \ii\epsilon  \right]^2\left[ \omega - E_{kp} - s_1 E_{k} + \ii\epsilon \right]^2}\nn\\
&&=PV\left( \frac{1}{\left[E_{kp}^2-(\omega-s_1 E_k)^2\right]^2} \right) +
\frac{\ii\pi}{4 E_{kp}^2}\Bigg[ \delta'\left( \omega + E_{kp} - s_1 E_k  \right) + \delta'\left( \omega - E_{kp} - s_1 E_k  \right)\nn\\ 
&&-\frac{1}{E_{kp}}\delta\left( \omega + E_{kp} - s_1 E_k  \right)
+ \frac{1}{E_{kp}}\delta\left( \omega - E_{kp} - s_1 E_k  \right) 
\Bigg]\nn\\
&&=PV\left( \frac{1}{\left[E_{kp}^2-(\omega-s_1 E_k)^2\right]^2} \right) +
\frac{\ii\pi}{4 E_{kp}^2}\sum_{s_2=\pm}\left[ 
\delta'\left( \omega - s_2 E_{kp} - s_1 E_k  \right) - \frac{1}{s_2 E_{kp}}\delta\left( \omega + s_2 E_{kp} - s_1 E_k  \right)
\right]
\eea
In a similar fashion:
\bea
&&\lim_{\epsilon\rightarrow 0}\frac{1}{\left[E_k^2-(\omega+\ii\epsilon+s_2 E_{kp})^2\right]^2}
=PV\left(\frac{1}{\left[E_k^2-(\omega+s_2 E_{kp})^2\right]^2}\right) +
\frac{\ii\pi}{4 E_{k}^2}\left[ \delta'\left( \omega + E_{k} + s_2 E_{kp}  \right) + \delta'\left( \omega - E_{k} + s_2 E_{kp}  \right)\right.\nn\\ 
&&\left.-\frac{1}{E_{k}}\delta\left( \omega +  E_{k} + s_2 E_{kp}  \right)
+ \frac{1}{E_{k}}\delta\left( \omega -  E_{k} + s_2 E_{kp}  \right) 
\right]\nn\\
&&=PV\left(\frac{1}{\left[E_k^2-(\omega+s_2 E_{kp})^2\right]^2}\right) +
\frac{\ii\pi}{4 E_{k}^2}\sum_{s_1=\pm}\left[ \delta'\left( \omega - s_1 E_{k} + s_2 E_{kp}  \right) + \frac{1}{s_1 E_{k}}\delta\left( \omega - s_1 E_{k} + s_2 E_{kp}  \right) \right]
\eea

For the square root:
\bea
&&\frac{1}{\sqrt{\left( \omega + \ii\epsilon - s_1 E_k  \right)^2 - m_f^2}}=\frac{1}{\sqrt{|(\omega-s_1 E_k)^2 - m_f^2|}\sqrt{1+2\ii\epsilon\frac{(\omega-s_1 E_k )}{(\omega-s_1 E_k)^2 - m_f^2}}}\nn\\
&&= \frac{1}{\sqrt{|(\omega-s_1 E_k)^2 - m_f^2|}\left(1 + \ii \epsilon \frac{(\omega-s_1 E_k )}{(\omega-s_1 E_k)^2 - m_f^2} \right)}\nn\\
&&= \frac{1}{\sqrt{|(\omega-s_1 E_k)^2 - m_f^2|}\left(\frac{\omega-s_1 E_k}{\left( \omega-s_1 E_k \right)^2-m_f^2} \right)\left(\frac{(\omega-s_1 E_k)^2 - m_f^2}{(\omega-s_1 E_k )} + \ii \epsilon  \right)}\nn\\
&&= \frac{1}{\sqrt{\left|\left( \omega  - s_1 E_k  \right)^2 - m_f^2\right|}} 
- \ii\pi\frac{\left( \omega - s_1 E_k\right)^2-m_f^2}{\left(\omega-s_1 E_k\right)\sqrt{\left|\left( \omega  - s_1 E_k  \right)^2 - m_f^2\right|}}
\delta\left( \frac{\left( \omega - s_1 E_k\right)^2-m_f^2}{\left(\omega-s_1 E_k\right)} \right)\nn\\
&=& \frac{1}{\sqrt{\left|\left( \omega  - s_1 E_k  \right)^2 - m_f^2\right|}},
\eea
where in the final step we applied the identity $x\delta(x) = 0$.

Therefore, to compute the yields, we only retain the imaginary part of $S_1$. The latter implies to compute the following integral:
\bea
\int\frac{dk_z}{2\pi}\text{Im}[S_1(k_z,p_z,\omega)]=\sum_{i=1}^{2}\sum_{s_1=\pm}\sum_{s_2=\pm}\mathcal{I}^{(s_1,s_2)}_i(p_z,\omega),
\eea
where
\bea
\mathcal{I}^{(s_1,s_2)}_1(p_z,\omega) &=& \frac{1}{16}\int_{-\infty}^{\infty} dk_z\frac{\left[\left( \omega - s_1 E_k \right)^2 - E_{kp}^2 + 2 m_f^2\right]n_\text{F}(\beta s_1 E_k)}{ s_1 E_k E_{kp}^2\sqrt{|\left( \omega  - s_1 E_k  \right)^2 - m_f^2}|}\left[ 
\delta'\left( \omega + s_2 E_{kp} - s_1 E_k \right)\right.\nn\\
&&\left.- \frac{1}{s_2 E_{kp}}\delta\left( \omega + s_2 E_{kp} - s_1 E_k \right)
\right]\nn\\
&=& -\frac{1}{16}\int_{-\infty}^{\infty} dk_z \frac{n_\text{F}(\beta s_1 E_k)}{s_1 E_k E_{kp}^2}\left[ \frac{\left( \omega - s_1 E_k \right)^2 - E_{kp}^2 + 2 m_f^2}{s_2 E_{kp}\sqrt{|\left( \omega  - s_1 E_k  \right)^2 - m_f^2}|}\right.\nn\\ 
&&\left.+
\frac{\partial}{\partial\omega}\frac{\left( \omega - s_1 E_k \right)^2 - E_{kp}^2 + 2 m_f^2}{\sqrt{|\left( \omega  - s_1 E_k  \right)^2 - m_f^2}|}
\right]\delta\left( \omega + s_2 E_{kp} - s_1 E_k \right)\\
\mathcal{I}_2^{(s_1,s_2)}(p_z,\omega)&=&\frac{1}{16}\int_{-\infty}^{\infty} dk_z\frac{\sigma_{s_2}(k_z,\omega)n_\text{F}(\beta s_2 E_{kp})}{s_2 E_{kp}^3 E_k^2\left(E_{kp}^2-m_f^2\right)^{3/2}}\left[ \delta'\left( \omega - s_1 E_{k} + s_2 E_{kp}  \right)\right.\nn\\
&&\left.+ \frac{1}{s_1 E_{k}}\delta\left( \omega - s_1 E_{k} + s_2 E_{kp}  \right) \right]\nn\\
&=& \frac{1}{16}\int_{-\infty}^{\infty} dk_z \frac{n_\text{F}(\beta s_2 E_{kp})}{s_2 E_{kp}^3 E_k^2\left(E_{kp}^2-m_f^2\right)^{3/2}}\left[ \frac{\sigma_{s_2}(k_z,\omega)}{s_1 E_k} -\frac{\partial}{\partial\omega}\sigma_{s_2}(k_z,\omega) \right]\delta\left( \omega - s_1 E_{k} + s_2 E_{kp}  \right)
\eea
In the last step, we applied the identity
\bea
\int dk_z f(k_z,\omega)\delta'(\omega - b) = - \int dk_z \frac{\partial}{\partial\omega}f(k_z,\omega)\delta(\omega - b).
\eea
to reduce the integral expressions to the form
\bea
\mathcal{I}^{(s_1,s_2)}_1(p_z,\omega) &=& -\frac{1}{16}\int_{-\infty}^{\infty} dk_z \frac{n_\text{F}(\beta s_1 E_k)}{s_1 E_k E_{kp}^2}\left[ \frac{\left( \omega - s_1 E_k \right)^2 - E_{kp}^2 + 2 m_f^2}{s_2 E_{kp}\sqrt{\left( \omega  - s_1 E_k  \right)^2 - m_f^2}}\right.\nn\\ 
&&\left.+
\frac{\partial}{\partial\omega}\frac{\left( \omega - s_1 E_k \right)^2 - E_{kp}^2 + 2 m_f^2}{\sqrt{\left( \omega  - s_1 E_k  \right)^2 - m_f^2}}
\right]\delta\left( \omega + s_2 E_{kp} - s_1 E_k \right)\nn\\
&=& -\frac{1}{16}\Theta\left( \omega - \sqrt{p_z^2 + 4 m_f^2} \right)\sum_{s=\pm}\left.\frac{n_\text{F}(\beta s_1 E_k)}{s_1 E_k E_{kp}^2\left|\frac{s_1 k_z}{E_k}-\frac{s_2(k_z-p_z)}{E_{kp}} \right|}\left[ \frac{\left( \omega - s_1 E_k \right)^2 - E_{kp}^2 + 2 m_f^2}{s_2 E_{kp}\sqrt{\left( \omega  - s_1 E_k  \right)^2 - m_f^2}}\right.\right.\nn\\ 
&&\left.\left.+
\frac{\partial}{\partial\omega}\frac{\left( \omega - s_1 E_k \right)^2 - E_{kp}^2 + 2 m_f^2}{\sqrt{\left( \omega  - s_1 E_k  \right)^2 - m_f^2}}
\right]\right|_{k_z = k_s}\nn\\
&=& \frac{1}{8}\Theta\left( \omega - \sqrt{p_z^2 + 4 m_f^2} \right)\sum_{s=\pm}\left.\frac{\left(E_{kp}^4-3 m_f^2 E_{kp}^2 + m_f^4\right) n_F\left(\beta s_1 E_k \right) }{s_1 E_k s_2 E_{kp}^3\left( E_{kp}^2 - m_f^2 \right)^{3/2}\left|\frac{s_1 k_z}{E_k}-\frac{s_2(k_z-p_z)}{E_{kp}} \right|}\right|_{k_z = k_s}
\eea
In order to compute the integrals $\mathcal{I}_i^{(s_1,s_2)}$, the delta function is handled with the same method described in Eq.~\eqref{eq:diraccompuesta}--\eqref{eq:E1E2}, assuming $\omega > 0$ which implies $|\left( \omega  - s_1 E_k  \right)^2 - m_f^2| =  \left(\omega  - s_1 E_k  \right)^2 - m_f^2$, as follows
\bea
\mathcal{I}^{(s_1,s_2)}_2(p_z,\omega) &=& \frac{1}{16}\int_{-\infty}^{\infty} dk_z \frac{n_\text{F}(\beta s_2 E_{kp})}{s_2 E_{kp}^3 E_k^2\left(E_{kp}^2-m_f^2\right)^{3/2}}\left[ \frac{\sigma_{s_2}(k_z,\omega)}{s_1 E_k} -\frac{\partial}{\partial\omega}\sigma_{s_2}(k_z,\omega) \right]\delta\left( \omega - s_1 E_{k} + s_2 E_{kp}  \right)\nn\\
&=& \frac{1}{16} \Theta\left( \omega - \sqrt{p_z^2 + 4 m_f^2} \right)\sum_{s=\pm}\left.\frac{n_\text{F}(\beta s_2 E_{kp})}{s_2 E_{kp}^3 E_k^2\left(E_{kp}^2-m_f^2\right)^{3/2}\left|\frac{s_1 k_z}{E_k}-\frac{s_2(k_z-p_z)}{E_{kp}} \right|}\left[ \frac{\sigma_{s_2}(k_z,\omega)}{s_1 E_k}\right.\right.\nn\\
&&\left.\left.-\frac{\partial}{\partial\omega}\sigma_{s_2}(k_z,\omega) \right]\right|_{k_z = k_s}\nn\\
&=& -\frac{1}{8}\Theta\left( \omega - \sqrt{p_z^2 + 4 m_f^2} \right)\sum_{s=\pm}\left.\frac{\left(E_{kp}^4-3 m_f^2 E_{kp}^2 + m_f^4\right) n_F\left(\beta s_2 E_{kp} \right) }{s_1 E_k s_2 E_{kp}^3\left( E_{kp}^2 - m_f^2 \right)^{3/2}\left|\frac{s_1 k_z}{E_k}-\frac{s_2(k_z-p_z)}{E_{kp}} \right|}\right|_{k_z = k_s}.
\eea

Therefore:
\bea
\text{Im}\left[g_{\mu\nu}\tau_1^{\mu\nu}\right]&=&-\frac{8m_f^2\qB}{\pi}\exp\left(-\frac{2\pt^2}{3\qB}\right)\sum_{s_1=\pm}\sum_{s_2=\pm}\sum_{i=1}^{2}\mathcal{I}^{(s_1,s_2)}_i(p_z,\omega)\nn\\
&=& \frac{m_f^2\qB}{\pi}\Theta\left( \omega - \sqrt{p_z^2 + 4 m_f^2} \right)\exp\left(-\frac{2\pt^2}{3\qB}\right)
\sum_{s_1=\pm}\sum_{s_2=\pm}\sum_{s=\pm}\mathcal{I}^{(s_1,s_2)}(p_z,\omega,s),
\label{ImgT3}
\eea
where we defined
\bea
&&\mathcal{I}^{(s_1,s_2)}(p_z,\omega,s)\equiv -\left.\frac{\left(E_{kp}^4-3 m_f^2 E_{kp}^2 + m_f^4\right) \left[n_F\left(\beta s_2 E_{kp} \right) - n_F\left(\beta s_1 E_{k} \right) \right]}{s_1 E_k s_2 E_{kp}^3\left( E_{kp}^2 - m_f^2 \right)^{3/2}\left|\frac{s_1 k_z}{E_k}-\frac{s_2(k_z-p_z)}{E_{kp}} \right|}\right|_{k_z = k_s}
\eea

\subsection{Computation of $g_{\mu\nu}\tau_3^{\mu\nu}$}

After integrating the perpendicular momenta, $\tau^{\mu\nu}_3$ takes the form:
\bea
\tau_3^{\mu\nu} = \frac{48\ii\pi\qB}{(2\pi)^2} \exp\left(-\frac{2\pt^2}{3\qB}\right)\int\frac{d^2k_\parallel}{(2\pi)^2}\frac{(m_f^2+\kp\cdot(\pp-\kp))(g_\parallel^{\mu\nu}-g_\perp^{\mu\nu})+(\kp^\mu - \pp^\mu)\kp^\nu+ \kp^\mu(\kp^\nu - \pp^\nu)}{(\kp^2-m_f^2)\left[(k-p)_\parallel^2-m_f^2\right]\sqrt{(k_0-p_0)^2-m_f^2}},
\eea
so that
\bea
g_{\mu\nu}\tau_3^{\mu\nu}=\frac{96\ii\pi\qB}{(2\pi)^2} \exp\left(-\frac{2\pt^2}{3\qB}\right)\int\frac{d^2k_\parallel}{(2\pi)^2}\frac{\kp\cdot(\kp-\pp)}{(\kp^2-m_f^2)\left[(k-p)_\parallel^2-m_f^2\right]\sqrt{(k_0-p_0)^2-m_f^2}}.
\eea

As in previous cases, we perform a Wick rotation onto Euclidean space, in such a way that the expression takes the form:
{\small
\bea
g_{\mu\nu}\tau_3^{\mu\nu}=-\frac{96\pi\qB}{(2\pi)^2} e^{-\frac{2\pt^2}{3\qB}}\int\frac{dk_z}{2\pi}T\sum_{n=-\infty}^{+\infty}\frac{1}{\sqrt{(\ii\omega_n-\ii\nu_l)^2-m_f^2}}\frac{(\ii\wn)(\ii\omega_n-\ii\nu_l)-k_z(k_z-p_z)}{\left[(\ii\omega_n)^2-k_z^2-m_f^2\right]\left[(\ii\omega_n-\ii\nu_l)^2-(k_z-p_z)^2-m_f^2\right]}.\nn\\
\eea
}

Let us define the infinite Matsubara sum
\bea
S_2=T\sum_{n=-\infty}^{+\infty}\frac{1}{\sqrt{(\ii\omega_n-\ii\nu_l)^2-m_f^2}}\frac{(\ii\wn)(\ii\omega_n-\ii\nu_l)-k_z(k_z-p_z)}{\left[(\ii\omega_n)^2-k_z^2-m_f^2\right]\left[(\ii\omega_n-\ii\nu_l)^2-(k_z-p_z)^2-m_f^2\right]}\equiv \sum_{n=-\infty}^{+\infty}f_2(\ii\omega_n).
\eea

In order to apply the same procedure of the computation of $g_{\mu\nu}\tau^{\mu\nu}_1$, we use the same contourn of Fig.~\ref{fig:contorno_suma_1}, so that
\bea
\frac{1}{2\pi\ii}\oint dz g(z)f_2(z)&=&\sum_{z_q    =\pm z_0,z_1^{\pm}}\left.\text{Res}\left\{g(z)f_2(z)\right\}\right|_{z_q}\nn\\
&=&g(z_0)\text{Res} [f_2(z)]_{z_0}+g(-z_0)\text{Res} [f_2(z)]_{-z_0}+g(z_1^+)\text{Res} [f_2(z)]_{z_1^+}+g(z_1^-)\text{Res} [f_2(z)]_{z_1^-}.
\eea

Given that the branch points are the same in both cases, it is straightforward to find that:
   \bea
S_2=-\sum\text{Res}\left\{g(z)f_2(z)\right\}+\frac{T}{\pi}\text{PV}\int_{-m_f}^{m_f}dx\frac{\left[x(x+\ii\nu_l)-k_z(k_z-p_z)\right]g(x)}{\left[(x+\ii\nu_l)^2-(k_z^2+m_f^2)\right]\left[x^2-(k_z-p_z)^2-m_f^2\right]\sqrt{x^2-m_f^2}}.
\eea 

After the analytic continuation $\ii\nu_l\to\omega+\ii\epsilon$, the sum over residues takes the form:
\bea
\sum_{z_q    =\pm z_0,z_1^{\pm}}\left.\text{Res}\left\{g(z)f_2(z)\right\}\right|_{z_q}&=&-\sum_{s_1=\pm}\frac{\left[s_1 E_k(s_1 E_k-\omega)-k_z(k_z-p_z)\right]n_\text{F}(s_1\beta E_k)}{2 s_1 E_k\left[(\omega+\ii\epsilon- s_1 E_k)^2-E_{kp}^2\right]\sqrt{(\omega+\ii\epsilon-s_1 E_k)^2-m_f^2}}\nn\\
&&-\sum_{s_2=\pm}\frac{\left[s_2 E_{kp}(\omega+s_2 E_{kp})-k_z(k_z-p_z)\right]n_\text{F}(s_2\beta E_{kp})}{2s_2 E_{kp}\left[(\omega+\ii\epsilon+s_2 E_{kp})^2-E_k^2\right]\sqrt{E_{kp}^2-m_f^2}}
\eea

In order to take the limit $\epsilon\to 0$, we use Eq.~\eqref{PVquotient} which implies that:
\bea
-\frac{1}{\left( \omega + \ii \epsilon - s_1 E_k \right)^2 - E_{kp}^2 }&=&\frac{1}{\left[E_{kp}+\left( \omega + \ii \epsilon - s_1 E_k \right) \right]\left[ E_{kp}-\left( \omega + \ii \epsilon - s_1 E_k \right) \right]}\nn\\
&=&\frac{1}{2E_{kp}}\left[\frac{1}{\omega-s_1 E_k+E_{kp}+\ii\epsilon}-\frac{1}{ \omega -s_1 E_k -E_{kp}+\ii \epsilon}\right]\nn\\
&=&\frac{1}{2E_{kp}}PV\left[\frac{1}{\omega-s_1 E_k+E_{kp}}-\frac{1}{ \omega -s_1 E_k -E_{kp}}\right]\nn\\
&-&\frac{\ii\pi}{2E_{kp}}\left[\delta(\omega-s_1 E_k+E_{kp})-\delta(\omega -s_1 E_k-E_{kp})\right]\nn\\
&=& PV\sum_{s_2=\pm}\frac{1}{2 s_2 E_{kp}}\frac{1}{\omega-s_1 E_k+ s_2 E_{kp}} - \ii \pi\sum_{s_2=\pm}\frac{1}{2 s_2 E_{kp}}\delta\left( \omega-s_1 E_k+ s_2 E_{kp} \right)
\eea
\bea
-\frac{1}{\left( \omega + \ii\epsilon+ s_2 E_{kp}\right)^2 - E_k^2}&=&-PV\sum_{s_1=\pm}\frac{1}{2 s_1 E_{k}}\frac{1}{\omega+s_2 E_{kp}- s_1 E_{k}}+\ii\pi\sum_{s_1=\pm}\frac{1}{2 s_1 E_k}\delta(\omega+s_2 E_{kp}-s_1 E_{k}).
\eea

On the other hand:
\bea
&&\frac{1}{\sqrt{\left( \omega + \ii\epsilon - s_1 E_k  \right)^2 - m_f^2}}=\frac{1}{\sqrt{|(\omega-s_1 E_k)^2 - m_f^2|}\sqrt{1+2\ii\epsilon\frac{(\omega-s_1 E_k )}{(\omega-s_1 E_k)^2 - m_f^2}}}\nn\\
&&= \frac{1}{\sqrt{|(\omega-s_1 E_k)^2 - m_f^2|}\left(1 + \ii \epsilon \frac{(\omega-s_! E_k )}{(\omega-s_1 E_k)^2 - m_f^2} \right)}\nn\\
&&= \frac{1}{\sqrt{|(\omega-s_1 E_k)^2 - m_f^2|}\left(\frac{\omega-s_! E_k}{\left( \omega-s_1 E_k \right)^2-m_f^2} \right)\left(\frac{(\omega-s_1 E_k)^2 - m_f^2}{(\omega-s_1 E_k )} + \ii \epsilon  \right)}\nn\\
&&= \frac{1}{\sqrt{\left|\left( \omega  - s_1 E_k  \right)^2 - m_f^2\right|}} 
- \ii\pi\frac{\left( \omega - s_1 E_k\right)^2-m_f^2}{\left(\omega-s_1 E_k\right)\sqrt{\left|\left( \omega  - s_1 E_k  \right)^2 - m_f^2\right|}}
\delta\left( \frac{\left( \omega - s_1 E_k\right)^2-m_f^2}{\left(\omega-s_1 E_k\right)} \right)\nn\\
&=& \frac{1}{\sqrt{\left|\left( \omega  - s_1 E_k  \right)^2 - m_f^2\right|}}
\eea
where in the final step we applied the identity $x\delta(x) = 0$.

Therefore, to compute the yields, we only retain the imaginary part of $S_2$. The latter implies to compute the following integral:
\bea
\int\frac{dk_z}{2\pi}\text{Im}[S_2(k_z,p_z,\omega)]=\sum_{s_1=\pm }\sum_{s_2=\pm}\sum_{i=1}^2\mathcal{J}^{(s_1,s_2)}_i(p_z,\omega),
\eea
so that
\bea
\text{Im}\left[g_{\mu\nu}\tau_3^{\mu\nu}\right]=-\frac{24\qB}{\pi}\exp\left(-\frac{2\pt^2}{3\qB}\right)\sum_{s_1=\pm }\sum_{s_2=\pm}\sum_{i=1}^2\mathcal{J}^{(s_1,s_2)}_i(p_z,\omega),
\label{ImgT3}
\eea
where
\begin{subequations}
\bea
\mathcal{J}_1^{(s_1,s_2)}(p_z,\omega)&=&-\frac{1}{8}\int dk_z\frac{\left[s_1 E_k \left(s_1 E_k - \omega \right)-k_z(k_z-p_z)\right]n_\text{F}(s_1\beta E_k)}{s_1 s_2 E_k E_{kp}\sqrt{\left|\left(\omega - s_1 E_k \right)^2 - m_f^2\right|}}\delta(\omega-s_1 E_k + s_2 E_{kp})
\eea
\bea
\mathcal{J}_2^{(s_1,s_2)}(p_z,\omega)=\frac{1}{8}\int dk_z\frac{\left[s_2 E_{kp}(\omega+s_2 E_{kp})-k_z(k_z-p_z)\right] n_\text{F}(s_2 \beta E_{kp})}{s_1 s_2 E_k E_{kp}\sqrt{E_{kp}^2-m_f^2}}\delta(\omega-s_1E_k +s_2 E_{kp})
\eea
\end{subequations}

With the method described in Eq.~\eqref{eq:diraccompuesta}--Eq.~\eqref{eq:E1E2}, the delta function is handled to compute the integrals above, to arrive at the explicit formulae
\bea
\mathcal{J}_1^{(s_1,s_2)}(p_z,\omega)
&=& -\frac{1}{8}\Theta\left(\omega -\sqrt{p_z^2 + 4 m_f^2} \right)\sum_{s=\pm} \left.\frac{\left[s_1 E_k s_2 E_{kp}-k_z(k_z-p_z)\right]n_\text{F}(\beta s_1 E_k)}{s_1 s_2 E_k E_{kp}\sqrt{E_{kp}^2 - m_f^2}\left|\frac{ s_1  k_s}{E_k}-\frac{s_2 (k_s-p_z)}{E_{kp}} \right|}\right|_{k_z = k_s}
\eea

\bea
\mathcal{J}_2^{(s_1,s_2)}(p_z,\omega)
&=& \frac{1}{8}\Theta\left(\omega -\sqrt{p_z^2 + 4 m_f^2} \right)\sum_{s=\pm} \left.\frac{\left[s_2 E_{kp}s_1 E_k-k_z(k_z-p_z)\right]n_\text{F}(\beta s_2 E_{kp})}{s_1 s_2 E_k E_{kp}\sqrt{E_{kp}^2 - m_f^2}\left|\frac{s_1 k_s}{E_k}-\frac{s_2 (k_s-p_z)}{E_{kp}} \right|}\right|_{k_z = k_s}
\eea

Therefore, we finally obtain
\bea
\text{Im}\left[g_{\mu\nu}\tau_3^{\mu\nu}\right]&=&-\frac{24\qB}{\pi}\exp\left(-\frac{2\pt^2}{3\qB}\right)\sum_{s_1=\pm}\sum_{s_2=\pm}\sum_{i=1}^{2}\mathcal{J}_i^{(s_1,s_2)}(p_z,\omega)\nn\\
&=& -\frac{3}{\pi}|q_f B|\Theta\left(\omega -\sqrt{p_z^2 + 4 m_f^2} \right)\exp\left(-\frac{2\pt^2}{3\qB}\right)\sum_{s_1=\pm}\sum_{s_2=\pm}\sum_{s=\pm}\mathcal{J}^{(s_1,s_2)}(p_z,\omega,s)
\label{ImgfT3}
\eea
where we defined
\bea
\mathcal{J}^{(s_1,s_2)}(p_z,\omega,s) &\equiv& \left.\frac{\left[s_1 E_k s_2 E_{kp}-k_s(k_s-p_z)\right]\left\{n_\text{F}(\beta s_2 E_{kp})-n_\text{F}(\beta s_1 E_k)\right\}}{s_1 s_2 E_k E_{kp}\sqrt{E_{kp}^2 - m_f^2}\left|\frac{ s_1  k_s}{E_k}-\frac{s_2 (k_s-p_z)}{E_{kp}} \right|}\right|_{k_z = k_s}
\eea
where the $k_s$ are given by Eq.~\eqref{roots1}. Clearly, if $\omega-s_1E_k+s_2E_{kp}\neq0$, the integral vanishes. Finally, by recognizing that the last expression depends only on the two energy values $E_+^{(s)} = s_1E_k(k_s)$ and 
$E_-^{(s)}=s_2E_{kp}(k_s)$ as defined in Eq.~\eqref{eq:E1E2}, and identifying the obvious constraints $s_1 E_{+}^{(s)} = E_k(k_s)>0$ and $s_2 E_{-}(k_s) = E_{kp}(k_s)>0$, the previous expression can be written in the final form quoted in the main text, Eq.~\eqref{eq:IJ} and Eq.~\eqref{eq:RDelta}.

\end{document}